\documentclass[aps,twocolumn,floats,floatfix,showpacs,amssymb,prd,superscriptaddress,nofootinbib]{revtex4-1}

\usepackage[utf8]{inputenc}
\usepackage{graphicx,amssymb,amsmath,amsthm,amsfonts,epsfig,times,natbib}

\usepackage[usenames,dvipsnames]{color}
\usepackage{epstopdf}
\usepackage{amsmath,amssymb}
\usepackage{tensor}
\usepackage{mathtools}
\usepackage{amsbsy}
\usepackage{bm,url}
\usepackage{float}
\usepackage{graphicx}
\usepackage[linktocpage]{hyperref}
\usepackage[usenames]{color}
\usepackage{dcolumn}
\usepackage{latexsym}
\usepackage{rotating}
\usepackage{color}
\usepackage{longtable}
\usepackage{enumerate}
\usepackage{booktabs}
\usepackage[utf8]{inputenc}
\definecolor{oxfordblue}{rgb}{0.0, 0.13, 0.28}
\definecolor{burgundy}{rgb}{0.5, 0.0, 0.13}
\definecolor{darkolivegreen}{rgb}{0.33, 0.42, 0.18}
\definecolor{darkblue}{rgb}{0,0,0.5}
\definecolor{richcarmine}{rgb}{0.84, 0.0, 0.25}
\definecolor{darkblue}{rgb}{0,0,0.5}
\definecolor{venetianred}{rgb}{0.78, 0.03, 0.08}
\definecolor{skobeloff}{rgb}{0.0, 0.48, 0.45}
\hypersetup{colorlinks=true, citecolor=darkblue, linkcolor=darkblue,
urlcolor = darkblue}

\def\nn{\nonumber}

\def\apjl{\rm{ApJ}}
\def\apjs{\rm{ApJS}}
\def\ao{\rm{Appl.~Opt.}}

\def\aap{\rm{A\&A}}

\newcommand{\ben}{\begin{enumerate}}
\newcommand{\een}{\end{enumerate}}

\def\be{\begin{equation}}
\def\ee{\end{equation}}
\def\bea{\begin{eqnarray}}
\def\eea{\end{eqnarray}}
\def\nn{\nonumber}
\newcommand{\beq}{\begin{eqnarray}}
\newcommand{\eeq}{\end{eqnarray}} 
\newcommand{\ba}{\begin{align}}
\newcommand{\ea}{\end{align}}

\begin{document}

\title{Black Hole Spectroscopy: Systematic Errors and Ringdown Energy Estimates}

\author{Vishal Baibhav}\email{vbaibhav@go.olemiss.edu}
\affiliation{Department of Physics and Astronomy, University of Mississippi,
University, Mississippi 38677, USA}
\author{Emanuele Berti}\email{eberti@olemiss.edu}
\affiliation{Department of Physics and Astronomy, University of Mississippi,
University, Mississippi 38677, USA}
\affiliation{CENTRA, Departamento de F\'{\i}sica, Instituto Superior
T\'ecnico, Universidade de Lisboa,
Avenida Rovisco Pais 1, 1049 Lisboa, Portugal}
\author{Vitor Cardoso}\email{vitor.cardoso@ist.utl.pt}
\affiliation{CENTRA, Departamento de F\'{\i}sica, Instituto Superior
T\'ecnico, Universidade de Lisboa,
Avenida Rovisco Pais 1, 1049 Lisboa, Portugal}
\affiliation{Perimeter Institute for Theoretical Physics, Waterloo, Ontario N2L 2Y5, Canada}
\author{Gaurav Khanna}\email{gkhanna@umassd.edu}
\affiliation{Department  of  Physics  \&  Center  for  Scientific  Computing  and  Visualization  Research,
University  of  Massachusetts  Dartmouth,  North  Dartmouth,  MA  02747,  USA}

\begin{abstract}
  The relaxation of a distorted black hole to its final state provides
  important tests of general relativity within the reach of current
  and upcoming gravitational wave facilities. In black hole
  perturbation theory, this phase consists of a simple linear
  superposition of exponentially damped sinusoids (the quasinormal
  modes) and of a power-law tail. How many quasinormal modes are
  necessary to describe waveforms with a prescribed precision? What
  error do we incur by only including quasinormal modes, and not
  tails? What other systematic effects are present in current
  state-of-the-art numerical waveforms? These issues, which are basic
  to testing fundamental physics with distorted black holes, have
  hardly been addressed in the literature. We use numerical relativity
  waveforms and accurate evolutions within black hole perturbation
  theory to provide some answers. We show that (i) a determination of
  the fundamental $l=m=2$ quasinormal mode to within $1\%$ or better
  requires the inclusion of at least the first overtone, and
  preferably of the first two or three overtones; (ii) a determination
  of the black hole mass and spin with precision better than $1\%$
  requires the inclusion of at least two quasinormal modes for any
  given angular harmonic mode $(\ell,\,m)$.  We also improve on
  previous estimates and fits for the ringdown energy radiated in the
  various multipoles. These results are important to quantify
  theoretical (as opposed to instrumental) limits in parameter
  estimation accuracy and tests of general relativity allowed by
  ringdown measurements with high signal-to-noise ratio gravitational
  wave detectors.
\end{abstract}

\maketitle

%%%%%%%%%%%%%%%%%%%%%%%%%%%%%%%%%%%%%%%%%%%%%%%%%%%%%%%%%%%%%%%%%%%%%%%%%%%%%
\section{Introduction}
%%%%%%%%%%%%%%%%%%%%%%%%%%%%%%%%%%%%%%%%%%%%%%%%%%%%%%%%%%%%%%%%%%%%%%%%%%%%%
%
The historic LIGO gravitational wave~(GW) detections of binary black
hole (BH)
mergers~\cite{Abbott:2016blz,Abbott:2016nmj,Abbott:2017vtc,Abbott:2017oio}
ushered in a new era in astronomy. The growing network of Earth-based
interferometers and the future space-based detector LISA will probe
the nature of compact objects and test general relativity (GR) in
unprecedented
ways~\cite{TheLIGOScientific:2016src,Yunes:2016jcc,Gair:2012nm,Yunes:2013dva,Berti:2015itd}.
One of the most interesting prospects is the possibility to use GW
observations to measure the quasinormal mode (QNM) oscillation
frequencies of binary BH merger remnants. In GR, these oscillation
frequencies depend only on the remnant BH mass and spin, so these
measurements can identify Kerr BHs just like atomic spectra identify
atomic elements. This idea is often referred to as ``BH
spectroscopy''~\cite{Detweiler:1980gk,Dreyer:2003bv,Berti:2005ys,Berti:2009kk}).
In the context of modified theories of gravity, QNM frequencies
would inform us on possible corrections to GR and allow to constrain
specific theories~\cite{Blazquez-Salcedo:2016enn,Glampedakis:2017dvb}.
In other words, the payoff of BH spectroscopy is significant not only
as a tool to test GR~\cite{Cardoso:2016ryw,Cardoso:2017njb}, but also
as a tool to quantify the presence of event horizons in the spacetime
(by looking, for instance, for ``echoes'' in the relaxation
stage~\cite{Cardoso:2016rao,Cardoso:2016oxy,Abedi:2016hgu,Mark:2017dnq}).

In practice, there are two main obstacles to measuring multiple QNM
frequencies (i.e., to identifying multiple spectral lines). The first
is of a technological nature, and relates to the fact that rather
large signal-to-noise ratios (SNRs) are
required~\cite{Berti:2007zu}. Recent estimates suggest that most {\em
  individual} binary BH mergers detected by LISA could be used to do
BH spectroscopy, but significant technological improvements are
necessary for Earth-based detectors to achieve the necessary
SNR~\cite{Berti:2016lat,Bhagwat:2016ntk}.
However the sensitivity of upcoming detectors is constantly improving,
and there are good reasons to believe that this issue will eventually
be resolved.
The second challenge concerns systematic effects which might be
unaccounted for in our current theoretical or numerical understanding
of the waveforms. For example, it is well known that (even at the
level of linearized perturbation theory) the late-time decay of BH
fluctuations is not exponential but
polynomial~\cite{Price:1971fb,Berti:2009kk}.  Thus, one must question
the validity of exponentially damped sinusoids as a description of the
late-time GW signal (see e.g. recent work by Thrane {\em et al.}, who
claimed that spectroscopy will not possible even in the infinite SNR
limit~\cite{Thrane:2017lqn}). When does the exponential (QNM) falloff
give way to the polynomial tail? Are nonlinearities important, and how
do they affect the simple linearized predictions?
 
There are very few studies of the accuracy achievable in extracting
QNM frequencies from numerical simulations. Some of these studies
pointed out that the accuracy of numerical waveforms may be limited by
gauge choices or wave extraction
techniques~\cite{Buonanno:2009qa,Zimmerman:2011dx}. Therefore we ask:
what is the systematic deviation between BH perturbation theory
predictions and the QNM frequencies extracted from numerical
simulations? In other words, what is the size of systematic errors in
the extraction of QNM frequencies from current state-of-the-art
numerical simulations?
These questions are of paramount importance for any claims about
independent BH mass and spin extraction using ringdown waveforms, and
for any ringdown-based tests of GR.
 
We address these questions using public catalogs of numerical
relativity simulations (focusing on the Simulating eXtreme Spacetimes
(SXS) Gravitational Waveform Database~\cite{Mroue:2013xna}), as well
as extreme mass-ratio waveforms produced using the Kerr time-domain
perturbative code written by one of
us~\cite{Sundararajan:2010sr,Zenginoglu:2011zz}.

One of the main results of our analysis, validating a multitude of
studies in the past decade or so, is that a ``pure ringdown'' stage
does not exist {\it per se}, detached from the rest of the
waveform. In other words, the full glory and complexity of GR must be
accounted for when extracting physics. Nevertheless, the notion of
ringdown can be useful in the context of simple, independent checks on
the physics. We have in mind, for instance, ringdown-based tests of
the no-hair theorem or constraints on modified theories of
gravity. Accurate models of the amplitude and phase of each QNM are
necessary to perform such tests. In fact, these quantities are also
crucial to alleviate the problem of low SNRs in individual events by
combining posterior probability densities from multiple
detections~\cite{Meidam:2014jpa} or via coherent
stacking~\cite{Yang:2017zxs}.  At the moment, our ability to do
coherent stacking is limited by the theoretical understanding of
ringdown: stacking requires phase alignment between different angular
components of the radiation, which can only be achieved through a
better understanding of the excitation and starting times of
QNMs~\cite{Andersson:1995zk,Nollert:1998ys,nollertthesis,Berti:2006wq,Zhang:2013ksa}.
Most early studies of QNM excitation relied on the evolution of simple
initial data (e.g. Gaussian wave packets) in the Kerr
background~\cite{Krivan:1997hc,Dorband:2006gg}. After the 2005
numerical relativity breakthrough, some authors investigated QNM
excitation in the merger of nonspinning
BHs~\cite{Buonanno:2006ui,Berti:2007fi,Kamaretsos:2011um,London:2014cma},
but to this day there is little published work on spinning mergers
(with the notable exception of Ref.~\cite{Kamaretsos:2012bs}).
In this work we use numerical relativity simulations to fit the energy
of the modes for spin-aligned binaries, thus alleviating some of the
difficulties inherent in stacking signals for BH spectroscopy.

%%%%%%%%%%%%%%%%%%%%%%%%%%%%%%%%%%%%%%%%%%%%%%%%%%%%%%%%%%%%%%%%%%%%%%%%%%%%%
\section{Systematic Errors in Extracting Quasinormal Mode Frequencies}
%%%%%%%%%%%%%%%%%%%%%%%%%%%%%%%%%%%%%%%%%%%%%%%%%%%%%%%%%%%%%%%%%%%%%%%%%%%%%
%
In the ringdown phase the radiation is a superposition of damped
sinusoids with complex frequencies $\omega^{\ell m n}$ parametrized by
three integers: the spin-weighted spheroidal harmonic indices
($\ell,\,m)$ and an ``overtone index'' $n$, which sorts the
frequencies by their decay time (the fundamental mode $n=0$ has the
smallest imaginary part and the longest decay time).
The complex Penrose scalar $\Psi_4$ (and the strain $h$) can be
expanded as
\be
r \Psi^{\ell m}_4 =
\Theta(t-t_0^{\ell m})
\sum_{n=1}^N B^{\ell mn} 
\exp\left[{\rm i}(\omega^{\ell mn} (t-t_0^{\ell m}) +\phi^{\ell mn})\right]\,. 
\label{eq:QNMFit1}
\ee
where $\Theta(x)$ is the Heaviside function,
$\omega^{\ell mn}=\omega^{\ell mn}_{\rm r}+{\rm i}\omega^{\ell
  mn}_{\rm i}$ and $t_0^{\ell m}$ is the so-called ``starting time''
of ringdown for the given $(\ell,\,m)$.
Early studies used least-squares fits to extract QNM frequencies from
nonspinning binary BH merger simulations~\cite{Buonanno:2006ui}. Other
fitting procedures were proposed, but yield very similar
results~\cite{Berti:2007dg,Berti:2007fi,Zimmerman:2011dx}.  Therefore,
for simplicity, we will use a simple least-squares fit. For
illustration, we consider nonspinning SXS waveforms with mass ratios
$q=1$ (SXS:BBH:0180) and $q=3$ (SXS:BBH:0183), as well as waveforms
for point particles falling into a nonrotating BH. 

For point particle evolutions we fit the strain $h$.  When considering
the SXS comparable-mass merger waveforms we use the Penrose scalar, as
it is known to yield slightly better QNM
fits~\cite{Buonanno:2006ui,Buonanno:2009qa}, but we checked that our
main conclusions would remain valid had we used the strain $h$
instead. For the multipolar components $(\ell,\,m)=(2,\,2)$, $(3,\,3)$
and $(2,\,1)$, that usually dominate the radiation, we use waveforms
extrapolated to infinite extraction radius using a second-order
polynomial (as reported by the SXS collaboration, higher-order
polynomials could yield noisy results close to the merger). For the
$(4,\,4)$ and higher-order multipoles we found that the ringdown part
of the waveform does not converge with extraction radius for a large
number of simulations. Furthermore, the largest extraction radii
listed in the SXS catalog are different for different simulations, so
they cannot be compared directly.  We only used waveforms for which
the higher-order multipoles seem to converge, finding the EMOP energy
as a function of extraction radius, and then comparing all energies
(whether computed by interpolation or extrapolation) at an extraction
radius of $500 M$.

The fits are performed in two different ways in order to address
different aspects of the systematic error analysis:

\noindent
(i) {\em How accurately can we determine the ringdown frequencies
  themselves, without assuming any (no-hair theorem enforced) relation
  between the frequencies?} 

To answer this question we assume that
($\omega_{\rm r}^{\ell mn},\,\omega_{\rm i}^{\ell mn},\,B^{\ell
  mn},\,\phi^{\ell mn}$) in Eq.~(\ref{eq:QNMFit1}) are all unknown, so
we have a total of $4N$ fitting coefficients for an $N$-mode fit. Then
we look at the relative error between the real and imaginary part of
the fundamental QNM (as derived from the fit) and the predictions from
BH perturbation theory~\cite{Berti:2005ys,Berti:2009kk}. This fitting
procedure does not enforce the fact that, in GR, QNM frequencies are
uniquely determined by the BH mass and
spin~\cite{Berti:2005ys,Berti:2009kk}. Systematic errors computed in
this way can be seen as lower bounds on how much any given modified
theory must modify ringdown frequencies to be experimentally
resolvable from GR.

%%%%%%%%%%%% Plot QNM Fits: Error: all %%%%%%%%%%%%%%%%%%%
\begin{figure*}[htp]
  \includegraphics[width=\textwidth]{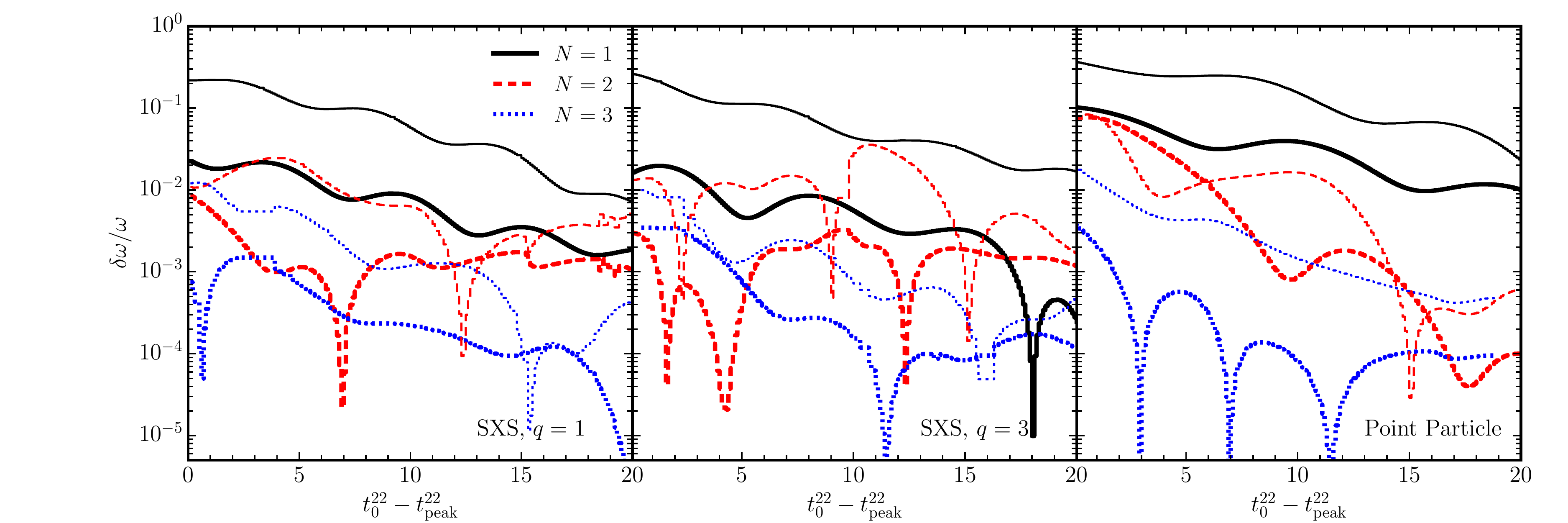}          
  \caption{\label{fig:QNMFitsErrFree} Fractional errors
    $\delta \omega_{\rm r}/\omega_{\rm r}$ (thick lines) and
    $\delta \omega_{\rm i}/\omega_{\rm i}$ (thin lines) between the
    fundamental $\ell=m=2$ QNM frequencies computed from BH
    perturbation theory and those obtained by fitting $N$ overtones to
    numerical waveforms according to method (i) (see text). Left: SXS
    waveforms, $q=1$; middle: SXS waveforms, $q=3$; right:
    point-particle waveforms.}
\end{figure*}
%%%%%%%%%%%%%%%%%%%%%%%%%%%%%%%%%%%%%%%%%%%%%%%%%%%%%%%

%%%%%%%%%%%% Plot QNM Fits: Error: all %%%%%%%%%%%%%%%%%%%
\begin{figure*}[htp]
  \includegraphics[width=\textwidth]{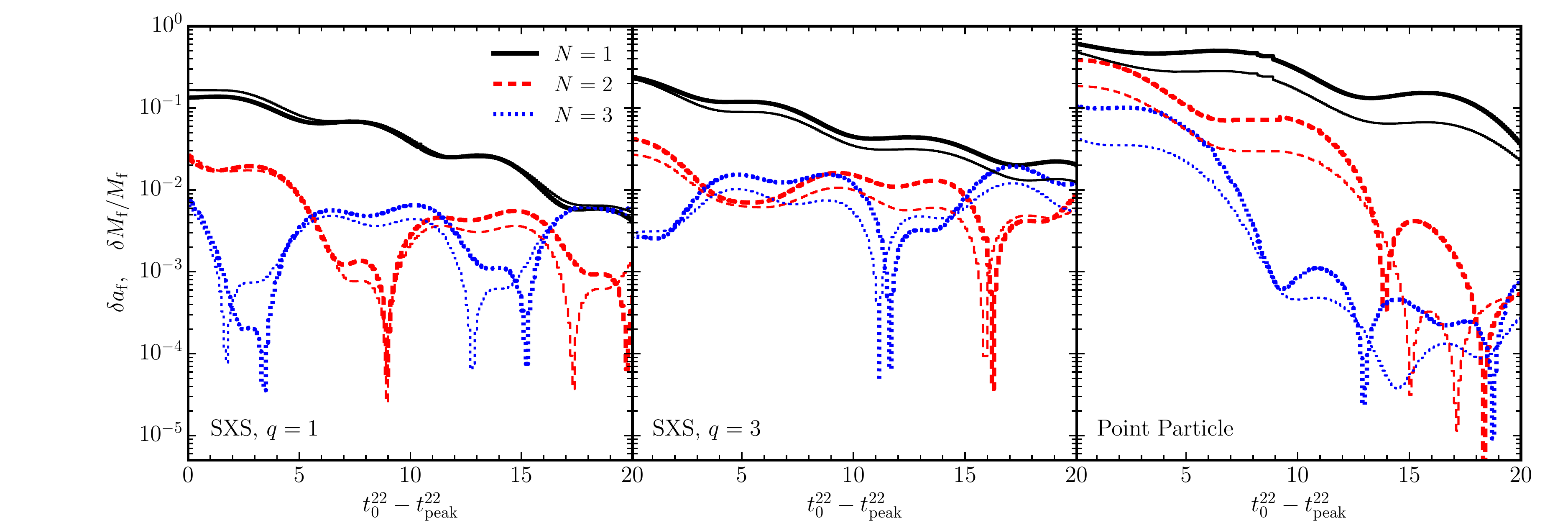}          
  \caption{\label{fig:QNMFitsErrMa} Error in the spin
    $\delta a_{\rm f}$ (thick lines) and fractional error in the mass
    $\delta M_{\rm f}/M_{\rm f}$ (thin lines) estimated by fitting $N$
    QNMs with $\ell=m=2$ according to method (ii) (see text). Left:
    SXS waveforms, $q=1$; middle: SXS waveforms, $q=3$; right:
    point-particle waveforms.}
\end{figure*}
%%%%%%%%%%%%%%%%%%%%%%%%%%%%%%%%%%%%%%%%%%%%%%%%%%%%%%%

The results are shown in Fig.~\ref{fig:QNMFitsErrFree}. BHs are poor
oscillators, so $\omega_{\rm r}$ is always easier to determine than
$\omega_{\rm i}$, and $\delta \omega_{\rm r}/\omega_{\rm r}$ is
typically an order of magnitude smaller than
$\delta \omega_{\rm i}/\omega_{\rm i}$.  Furthermore,
Fig.~\ref{fig:QNMFitsErrFree} shows that (contrary to the claims
of~\cite{Thrane:2017lqn}) adding overtones generally reduces the
systematic error in $\omega_{\rm r}$ and $\omega_{\rm i}$ for all mass
ratios. For SXS waveforms we found that including the $N=4$ mode would
not further improve the agreement, while for quasicircular inspirals
of point particles into nonrotating BHs
$\delta \omega_{\rm r}/\omega_{\rm r}$ and
$\delta \omega_{\rm i}/\omega_{\rm i}$ decreases to $\sim 10^{-4}$ and
$10^{-3}$, respectively.

\noindent
(ii) {\em How accurately can we determine the remnant's mass and spin
  from ringdown frequencies, assuming that GR is correct?} 

To answer this question we still consider
($B_{lm}^{(j)},\,\phi_{lm}^{(j)}$) as free parameters, but now we
enforce the condition that the QNM frequencies
$\omega_{\rm r,\,{\rm i}}^{\ell mn}$ must be functions of the remnant
BH mass $M_{\rm f}$ and dimensionless spin $a_{\rm f}$, so we have
only $2N+2$ fitting coefficients.
As shown in Fig.~\ref{fig:QNMFitsErrMa}, the accuracy in determining
both mass and spin is comparable to the accuracy in the poorest
determined quantity (i.e., $\omega_{\rm i}$). The trend is the same as
in Fig.~\ref{fig:QNMFitsErrFree}, and errors decrease as we include
more overtones.

The results in Figs.~\ref{fig:QNMFitsErrFree} and
\ref{fig:QNMFitsErrMa} disprove the claim of~\cite{Thrane:2017lqn}
that large-SNR detections cannot be used to perform BH spectroscopy,
but they also show that the relative error between quantities computed
in BH perturbation theory and those extracted from numerical
simulations currently saturates at $\sim 10^{-3}$. This ``saturation
effect'' is less problematic for the quasicircular inspiral of point
particles into Schwarzschild BHs, where relative errors can be reduced
by approximately one order of magnitude (we get worse agreement for
point particles falling into rotating BHs, where spherical-spheroidal
mode
mixing~\cite{Berti:2005gp,Buonanno:2006ui,Kelly:2012nd,Berti:2014fga}
must be taken into account).

This observation has an important implication: {\em further numerical
  or theoretical work is required to reduce systematic errors for
  comparable-mass binary BH mergers in the LISA band}, that may have
SNRs $\sim 10^3$ or higher~\cite{AmaroSeoane:2012km,Audley:2017drz}.

 %%%%%%%%%%%% Plot 33 in 22%%%%%%%%%%%%%%%%%%%
\begin{figure}%
\includegraphics[width=\columnwidth]{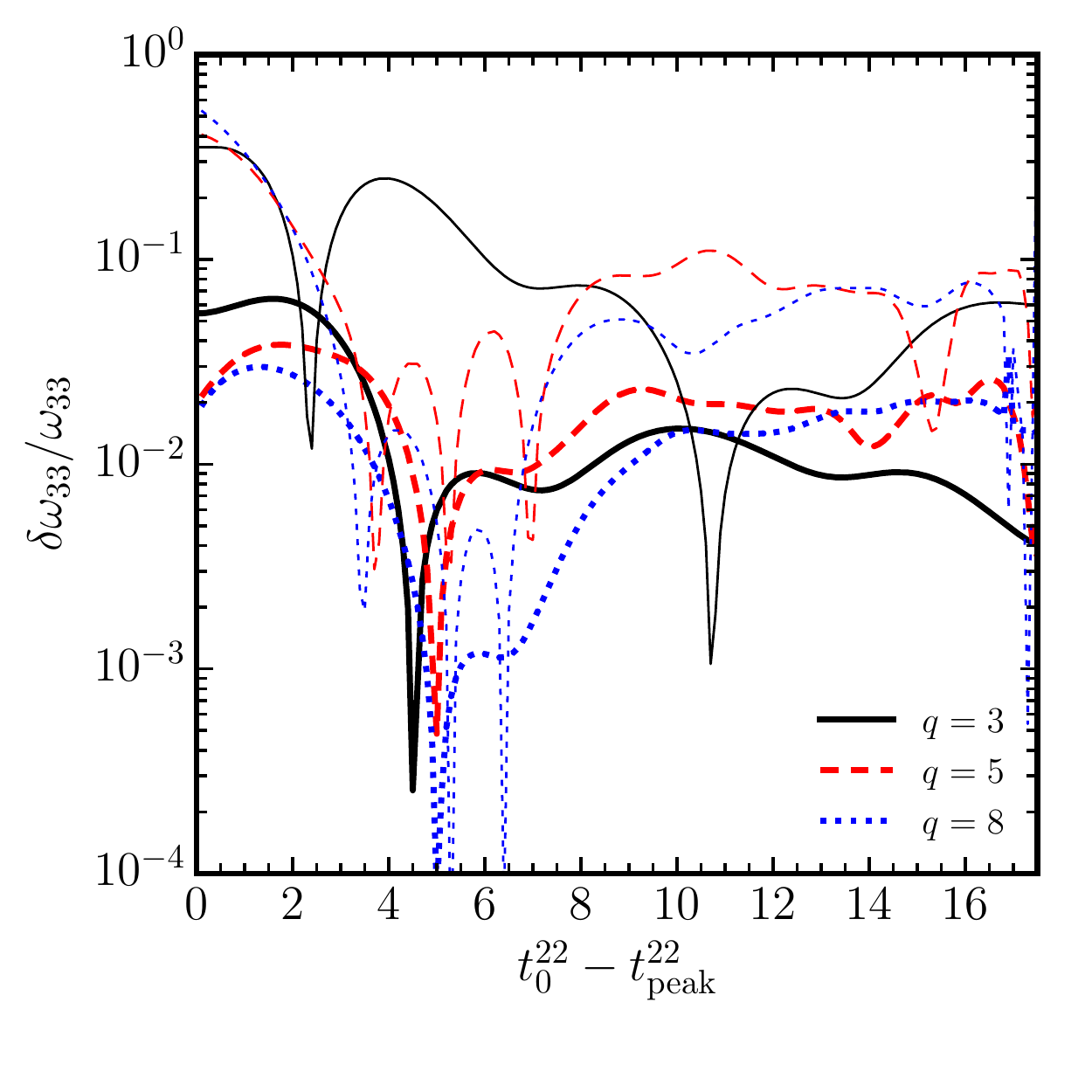}%
\caption{This figure shows how $(3,\,3)$ modes contaminate the
  $(2,\,2)$ components of unequal-mass BH mergers in the SXS
  waveforms. We fit the $(2,\,2)$ mode using a 3-mode fit and method
  (i) in the text. Then we plot the fractional errors
  $\delta \omega_{\rm r}/\omega_{\rm r}$ (thick lines) and
  $\delta \omega_{\rm i}/\omega_{\rm i}$ (thin lines) with respect to
  the fundamental $\ell = m = 3$ QNM frequencies from BH perturbation
  theory. This plot used the simulations labeled as SXS:BBH:0183 for
  $q=3$, SXS:BBH:0056 for $q=5$ and SXS:BBH:0063 for $q=8$.  }%
\label{fig:lm33}%
\end{figure}
%%%%%%%%%%%%%%%%%%%%%%%%%%%%%%%%%%%%%%%%%%%%%%%%%%%%%%%

The saturation discussed above may be related to an undesired feature
of SXS waveforms. It wass already noted in~\cite{Zimmerman:2011dx}
that the $\ell=m=2$ component of $\Psi_4$ in the SXS simulations
contains a spurious decaying mode corresponding to the fundamental
$\ell=m=4$ QNMs for $q=1$. We confirm their finding. Furthermore, as
we show in Fig.~\ref{fig:lm33}, a multi-mode fit of {\em
  unequal-mass} waveforms shows the presence of a spurious frequency
that matches quite well the fundamental QNM with $\ell=m=3$.

These spurious modes seem to be present only in the SXS
simulations. We did not find them in the public catalog of waveforms
from the Georgia Tech group~\cite{Jani:2016wkt}, nor in our own
point-particle waveforms.  Understanding the origin of these modes is
beyond the scope of this work. We speculate that they may be gauge or
wave extraction artifacts, but they are unlikely to come from
spherical-spheroidal mode mixing, which only mixes components with the
same $m$ and different
$\ell$'s~\cite{Berti:2005gp,Buonanno:2006ui,Kelly:2012nd,Berti:2014fga}.
Whatever their origin, these spurious modes must be understood if we
want to control systematics at the level required to do BH
spectroscopy with LISA.

%%%%%%%%%%%%%%%%%%%%%%%%%%%%%%%%%%%%%%%%%%%%%%%%%%%%%%%%%%%%%%%%%%%%%%%%%%%%%
\section{Ringdown Energies and Starting Times}
%%%%%%%%%%%%%%%%%%%%%%%%%%%%%%%%%%%%%%%%%%%%%%%%%%%%%%%%%%%%%%%%%%%%%%%%%%%%%
%
An important prerequisite to perform BH spectroscopy (whether via
single detections or by stacking) is to quantify the excitation of
QNMs, and to provide a definition of their starting times which is
suitable for data analysis purposes. Quite remarkably, we are aware of
only one paper that tried to quantify QNM excitation for spinning
binaries~\cite{Kamaretsos:2011um}. Here we improve on the results
of~\cite{Kamaretsos:2011um} by (i) using newer and more accurate
simulations from the SXS catalog, and (ii) implementing a better
criterion to determine {\em simultaneously} the energy (or relative
amplitude) of different ringdown modes, as well as their starting
times.

 %%%%%%%%%%%% Plot non spinning energy%%%%%%%%%%%%%%%%%%%
\begin{figure}%
\includegraphics[width=\columnwidth]{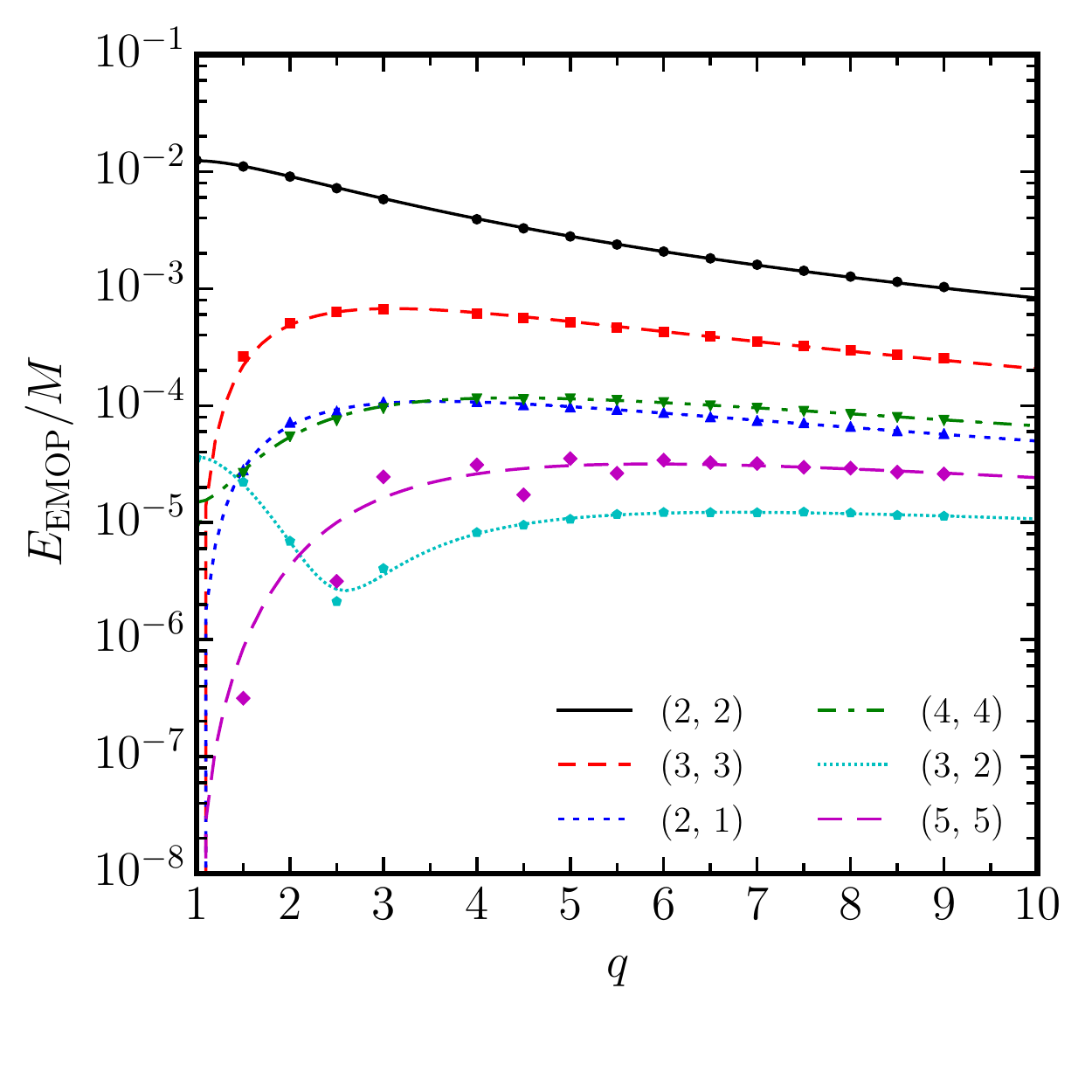}%
\caption{EMOP energies as a function of mass ratio for nonspinning
  binaries in the SXS catalog. The anomalous behavior of the $(3,\,2)$
  mode is due to spherical-spheroidal mode
  mixing~\cite{Berti:2005gp,Buonanno:2006ui,Kelly:2012nd,Berti:2014fga}:
  the contamination of the $(2,\,2)$ mode observed in the $(3,\,2)$
  mode is more prominent for comparable mass ratios.}%
\label{fig:noSpinE}%
\end{figure}
%%%%%%%%%%%%%%%%%%%%%%%%%%%%%%%%%%%%%%%%%%%%%%%%%%%%%%%

%%% Fitting Coefficients%%%%%%%%%%%%%%%%%%%%%%%
\newcolumntype{L}{>{$}l<{$}}
%% turns Tabular environment to Math Mode. (Mathematica gives the output as array (MathMode))

 \begin{table*}
   \caption{Fitting coefficients for the EMOP energy, along with the
     corresponding errors. A superscript ``0'' corresponds to the
     nonspinning contribution, while ``$s$'' denotes the spin-dependent
     contributions. Since poorly excited modes tend to be dominated by
     numerical noise, we have only considered modes with
     $E_\text{EMOP}\ge10^{-4} M$. We also dropped the $(4,\,4)$ mode
     data from some simulations where the EMOP energy did not converge
     as we increase the wave extraction radius.}
\label{tab:fittingCoeffETable}
\setlength\tabcolsep{9 pt}
\begin{tabular}{LLLLLLLLLLL}
\hline
  \text{Modes} & a^0 & b^0 & c^0 & a^s & b^s & c^s &
   d^s & e^s &\text{Max. Error} & \text{Mean Error}\\
  \hline
  \hline
 (2,\,2) & 0.303 & 0.571 & 0 & -0.07 & 0.255 &
   0.189 & -0.013 & 0.084 & 3.63\% & 0.64\% \\
 (3,\,3) & 0.157 & 0.671 & 0 & 0.163 & -0.187
   & 0.021 & 0.073 & 0 & 11.24\% & 2.32\%\\
 (2,\,1) & 0.099 & 0.06 & 0 & -0.067 & 0 & 0
   & 0 & 0 & 9.54\% & 2.01\% \\
 (4,\,4) & 0.122 & -0.188 & -0.964 & -0.207 &
   0.034 & -0.701 & 1.387 & 0.122 & 12.75\% & 1.93\% \\
 \hline
\end{tabular}
\end{table*}
%%%%%%%%%%%%%%%%%%%%%%%%%%%%%%%%%%%%%%%%%%%%%%%%

%%%%%%%%%%%% Plot Energy vs spin %%%%%%%%%%%%%%%%%%%
 \begin{figure*}[tp]
  	\begin{center}
  		\begin{tabular}{cc}
  			\epsfig{file=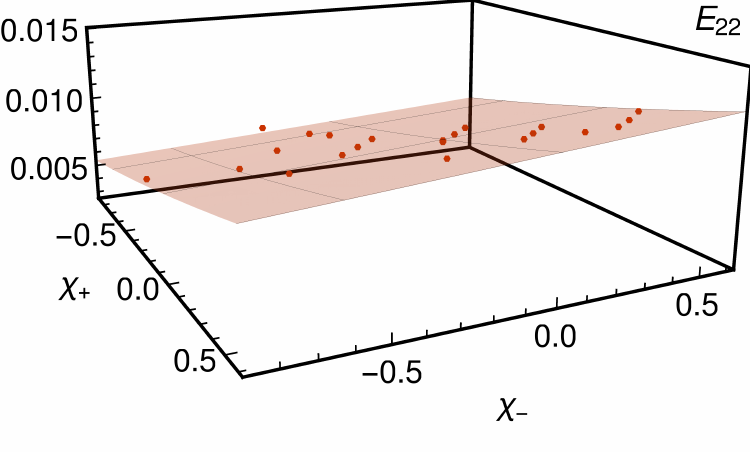, width=0.4 \linewidth}&
  			\epsfig{file=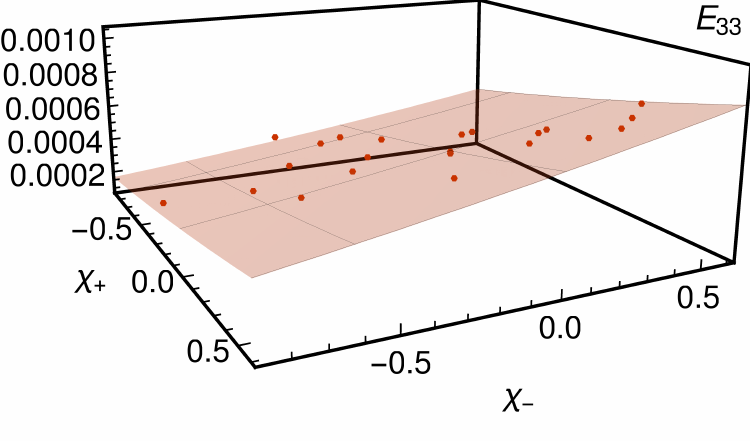, width=0.4 \linewidth}\\
  			\epsfig{file=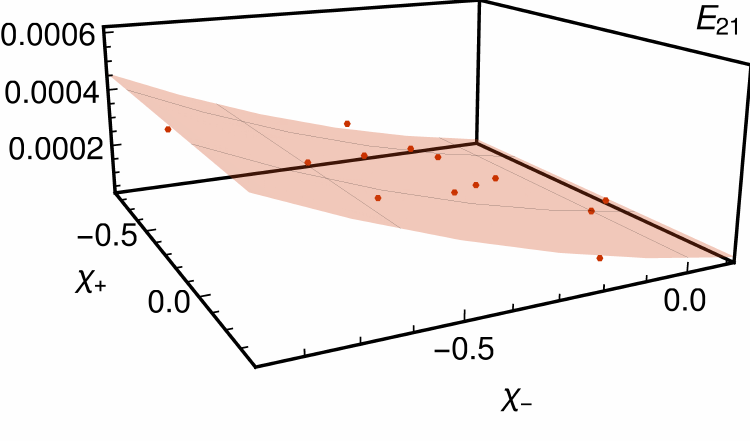, width=0.4 \linewidth}&
  			\epsfig{file=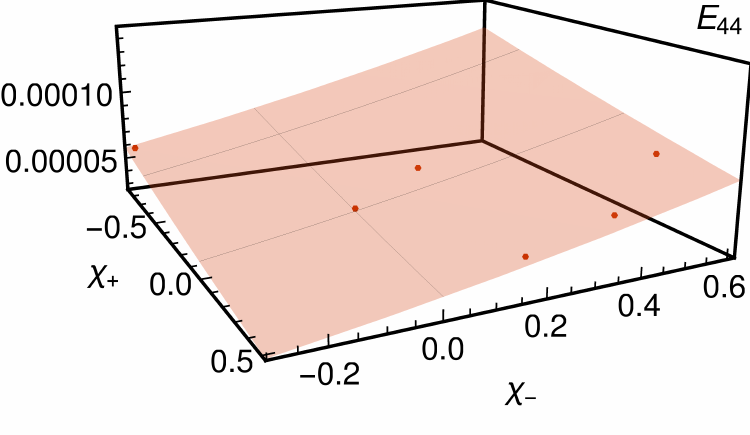, width=0.4 \linewidth}
  		\end{tabular}
  	\end{center}
  	\caption{EMOP energies $E_{\ell m}$ in different $(\ell,\,m)$
          modes for aligned-spin SXS simulations with $q=2$ as a
          function of $\chi_+$ and $\chi_-$, along with the fits given
          in Eq.~(\ref{eq:fit}).}
 \label{fig:EmopEnergiesS}
 \end{figure*}
 %%%%%%%%%%%%%%%%%%%%%%%%%%%%%%%%%%%%%%%%%%%%%%%%%%%%%%%
 
There is no unique, unambiguous way of defining such a starting time,
because ringdown is only an intermediate part of the full signal
resulting from the merger dynamics of the two-body system.
Nevertheless, a physically sensible, detector-independent criterion is
to decompose the full waveform into components ``parallel'' and
``perpendicular'' to the QNM. The ringdown starting time is defined as
the point where the energy ``parallel to the QNM'' is
maximized. Nollert, who introduced this concept, called this the
``energy maximized orthogonal projection'' (EMOP)~\cite{Nollert}. A
ringdown waveform starting at time $t_0$ has the form
\be
h_{\text{QNM}}=h^+_{\text{QNM}}+ih^{\times }_{\text{QNM}}\nn\\
=\Theta(t-t_0)\exp\left[{\rm i}(\omega t +\phi)\right]\,. 
\ee
Given the complex strain $h=h^++i h^\times$ from numerical relativity,
the energy ``parallel to the QNM'' $h_\text{QNM}$ is
\be
E_{\parallel }=
\frac{1}{8\pi}
\frac{\lvert\int_{t_0}\dot{h}
  \dot{h}^{*}_{{\rm QNM}}dt\rvert^2}
{\int_{t_0}
  \dot{h}_{{\rm QNM}} \dot{h}^*_{{\rm QNM}}dt}
=
\frac{\omega _{\rm i} \lvert\int_{t_0}\dot{h}
  \dot{h}^{*}_{{\rm QNM}}dt\rvert^2}
{4\pi\left(\omega _{\rm i}^2+\omega _{\rm r}^2\right)}
\,,
\label{eq:emopEnergyEq1}
\ee
where in the second equality we have explicitly evaluated the integral
in the denominator.
The ringdown starting time is defined as the lower limit of
integration $t_0$ such that $E_\parallel$ in
Eq.~(\ref{eq:emopEnergyEq1}) is maximum, and the EMOP energy is
$E_\text{EMOP}=\text{max}_{t_0}(E_\parallel)$.

Equation~(\ref{eq:emopEnergyEq1}) is an improvement over the
definition used in~\cite{Berti:2007fi}, where we first computed the
EMOP energy separately for the plus and cross polarizations, and then
averaged the starting time from the two polarizations.  Furthermore
$E_{\parallel}$ is independent of phase rotations in either the
numerical waveform ($h \to h e^{{\rm i}\theta}$) or in the QNM
($h_{\rm QNM} \to h_{\rm QNM} e^{{\rm i}\phi}$).
EMOP energies computed from the SXS waveforms for nonspinning binary
mergers are shown in Fig.~\ref{fig:noSpinE}. 

For binaries with aligned spins, a good fit to the EMOP energy in the
first few dominant $(\ell,\,m)$ modes is
\be
E_{\ell m}=\begin{cases}
\eta^2({\cal{A}}_{\ell m}^0+{\cal{A}}_{\ell m}^{\text{spin}})^2\,, &
\text{even}~m\,,\\
\eta^2(\sqrt{1-4
  \eta}{\cal{A}}_{\ell m}^0+{\cal{A}}_{\ell m}^{\text{spin}})^2\,, & 
\text{odd}~m\,,
\end{cases}\label{eq:fit}
\ee
where the nonspinning contribution ${\cal{A}}_{\ell m}^0$ is well
fitted by
\bea
&&{\cal{A}}_{\ell m}^0= a_{\ell m}^0 +b_{\ell m}^0 \eta\,, 
\quad (\ell,\,m)=(2,\,2),\,(3,\,3),\,(2,\,1)\,,
\nn
\\
&&{\cal{A}}_{\ell m}^0=a_{\ell m}^0+b_{\ell m}^0  \eta+ c_{\ell m}^0\eta^2\,, 
(\ell,\,m)=(3,\,2),\,(4,\,4),\,(5,\,5)\,,
\nn
\eea
and $\eta=q/(1+q)^2$ is the symmetric mass ratio. 
The contribution from the spins ${\cal{A}}_{\ell m}^\text{spin}$ can
be written in terms of the symmetric and asymmetric effective spins
\be
\chi_{\pm}\equiv \frac{m_1 \chi_1\pm m_2 \chi_2}{M}\,,
\ee
where $\chi_1$ and $\chi_2$ are the dimensionless spins of the two
BHs, and $\chi_+=\chi_{\rm eff}$ (the ``effective spin'' parameter
best measured by LIGO, which is conserved in post-Newtonian evolutions
at 2PN
order~\cite{Kesden:2010yp,Kesden:2014sla,Gerosa:2015tea,Gerosa:2015hba}).

We use the post-Newtonian inspired
fits~\cite{Barausse:2009xi,Pan:2010hz}
\begin{align}
{\cal{A}}_{22}^\text{spin}=&
\eta  \chi_+ \left(a_{22}^\text{s}+\frac{b_{22}^\text{s}}{q}+c_{22}^\text{s} q+d_{22}^\text{s} q^2\right)+e_{22}^\text{s} \delta \chi_-\,,
\nn\\
{\cal{A}}_{33}^\text{spin}=& 
\eta  \chi_- \left(a_{33}^\text{s}+\frac{b_{33}^\text{s}}{q}+c_{33}^\text{s} q\right)+d_{33}^\text{s} \delta  \chi_+
\,,
\nn\\
{\cal{A}}_{21}^\text{spin}=& a_{21}^\text{s} \chi_-\,,
\nn\\
{\cal{A}}_{44}^\text{spin}=&
\eta  \chi_+ \left(\frac{a_{44}^\text{s}}{q}+b_{44}^\text{s} q\right)+\delta  \eta  \chi_- \left(c_{44}^\text{s}+\frac{d_{44}^\text{s}}{q}+e_{44}^\text{s} q\right)
\,,
\label{eq:spinFit1-44}
\end{align}
where $\delta=\sqrt{1-4 \eta}=(q-1)/(q+1)$.  The fitting coefficients,
along with the mean and maximum percentage errors of each fit, are
listed in Table~\ref{tab:fittingCoeffETable}. The dependence of the
EMOP energy on spins is illustrated in Fig.~\ref{fig:EmopEnergiesS}
for simulations with mass ratio $q=2$.

%%%%%%%%%%%%%%%%%%%%%%%%%%%%%%%%%%%%%%%%%%%%%%%%%%%%%%%%%%%%%%%%%%%%%%%%%%%%%
\section{Conclusions}
%%%%%%%%%%%%%%%%%%%%%%%%%%%%%%%%%%%%%%%%%%%%%%%%%%%%%%%%%%%%%%%%%%%%%%%%%%%%%
The recent detection of gravitational waves by the LIGO/Virgo
collaboration makes the prospect of spectroscopic tests of general
relativity realistic in the near future.  
As detectors and data quality improve, a good understanding of the
ringdown stage will require an assessment of systematic errors
affecting the waveforms. Previous studies bounded environmental and
astrophysical effects in BH ringdown
waveforms~\cite{Barausse:2014tra}. In this work we started addressing
how numerical and/or theoretical limitations affect our ability to
perform BH spectroscopy. It is known that the late-time behavior of
any BH perturbation should be a power-law decay. Thus, a description
using exponentially damped sinusoids must eventually break down. 

We showed that no precise tests of GR nor any accurate measurement of
BH masses or spins are possible with single-mode templates: two or
three modes are necessary.

To facilitate spectroscopic tests (whether in single detections or via
stacking) we extended the EMOP calculations of
Ref.~\cite{Berti:2007fi} using the SXS waveforms in the case of
(anti-)aligned spins. In this preliminary study we neglected subtle
issues such as mode mixing, which is known to affect in particular the
$(3,\,2)$ mode~\cite{Buonanno:2006ui,Kelly:2012nd,Berti:2014fga}.
Further work is required to apply our results in gravitational-wave
data
analysis~\cite{Berti:2005ys,Berti:2007fi,Gossan:2011ha,Meidam:2014jpa,Ghosh:2016qgn}
or to understand how these systematics affect tests general relativity
with ringdown, e.g. within the ``post-Kerr'' framework proposed
in~\cite{Glampedakis:2017dvb}.

%%%%%%%%%%%%%%%%%%%%%%%%%%%%%%%%%%%%%%%%%%%%%%%%%%%%%%%%%%%%%%%%%%%%%%%%%%%%%
\noindent{\bf{\em Acknowledgments.}}
%%%%%%%%%%%%%%%%%%%%%%%%%%%%%%%%%%%%%%%%%%%%%%%%%%%%%%%%%%%%%%%%%%%%%%%%%%%%%
% \begin{acknowledgments}
We thank Bangalore Sathyaprakash, K.G. Arun and Daniel George for
useful discussions.
V.B. and E.B. are supported by NSF Grants No.~PHY-1607130
and AST-1716715, and by FCT contract IF/00797/2014/CP1214/CT0012 under
the IF2014 Programme.
V.C. acknowledges financial support provided under the European
Union's H2020 ERC Consolidator Grant ``Matter and strong-field
gravity: New frontiers in Einstein's theory'' grant agreement
no. MaGRaTh--646597. Research at Perimeter Institute is supported by
the Government of Canada through Industry Canada and by the Province
of Ontario through the Ministry of Economic Development $\&$
Innovation.
G.K. acknowledges research support from the National Science
Foundation (award No.~PHY-1701284) and Air Force Research Laboratory
(agreement No.~10-RI-CRADA-09).
This project has received funding from the European Union's Horizon
2020 research and innovation programme under the Marie
Sklodowska-Curie grant agreement No 690904.
The authors would like to acknowledge networking support by the COST
Action CA16104.

%\bibliography{EMOP}

\begin{thebibliography}{62}%
\makeatletter
\providecommand \@ifxundefined [1]{%
 \@ifx{#1\undefined}
}%
\providecommand \@ifnum [1]{%
 \ifnum #1\expandafter \@firstoftwo
 \else \expandafter \@secondoftwo
 \fi
}%
\providecommand \@ifx [1]{%
 \ifx #1\expandafter \@firstoftwo
 \else \expandafter \@secondoftwo
 \fi
}%
\providecommand \natexlab [1]{#1}%
\providecommand \enquote  [1]{``#1''}%
\providecommand \bibnamefont  [1]{#1}%
\providecommand \bibfnamefont [1]{#1}%
\providecommand \citenamefont [1]{#1}%
\providecommand \href@noop [0]{\@secondoftwo}%
\providecommand \href [0]{\begingroup \@sanitize@url \@href}%
\providecommand \@href[1]{\@@startlink{#1}\@@href}%
\providecommand \@@href[1]{\endgroup#1\@@endlink}%
\providecommand \@sanitize@url [0]{\catcode `\\12\catcode `\$12\catcode
  `\&12\catcode `\#12\catcode `\^12\catcode `\_12\catcode `\%12\relax}%
\providecommand \@@startlink[1]{}%
\providecommand \@@endlink[0]{}%
\providecommand \url  [0]{\begingroup\@sanitize@url \@url }%
\providecommand \@url [1]{\endgroup\@href {#1}{\urlprefix }}%
\providecommand \urlprefix  [0]{URL }%
\providecommand \Eprint [0]{\href }%
\providecommand \doibase [0]{http://dx.doi.org/}%
\providecommand \selectlanguage [0]{\@gobble}%
\providecommand \bibinfo  [0]{\@secondoftwo}%
\providecommand \bibfield  [0]{\@secondoftwo}%
\providecommand \translation [1]{[#1]}%
\providecommand \BibitemOpen [0]{}%
\providecommand \bibitemStop [0]{}%
\providecommand \bibitemNoStop [0]{.\EOS\space}%
\providecommand \EOS [0]{\spacefactor3000\relax}%
\providecommand \BibitemShut  [1]{\csname bibitem#1\endcsname}%
\let\auto@bib@innerbib\@empty
%</preamble>
\bibitem [{\citenamefont {Abbott}\ \emph
  {et~al.}(2016{\natexlab{a}})\citenamefont {Abbott} \emph
  {et~al.}}]{Abbott:2016blz}%
  \BibitemOpen
  \bibfield  {author} {\bibinfo {author} {\bibfnamefont {B.~P.}\ \bibnamefont
  {Abbott}} \emph {et~al.} (\bibinfo {collaboration} {Virgo, LIGO
  Scientific}),\ }\href {\doibase 10.1103/PhysRevLett.116.061102} {\bibfield
  {journal} {\bibinfo  {journal} {Phys. Rev. Lett.}\ }\textbf {\bibinfo
  {volume} {116}},\ \bibinfo {pages} {061102} (\bibinfo {year}
  {2016}{\natexlab{a}})},\ \Eprint {http://arxiv.org/abs/1602.03837}
  {arXiv:1602.03837 [gr-qc]} \BibitemShut {NoStop}%
%%CITATION = ARXIV:1602.03837;%%
\bibitem [{\citenamefont {Abbott}\ \emph
  {et~al.}(2016{\natexlab{b}})\citenamefont {Abbott} \emph
  {et~al.}}]{Abbott:2016nmj}%
  \BibitemOpen
  \bibfield  {author} {\bibinfo {author} {\bibfnamefont {B.~P.}\ \bibnamefont
  {Abbott}} \emph {et~al.} (\bibinfo {collaboration} {Virgo, LIGO
  Scientific}),\ }\href {\doibase 10.1103/PhysRevLett.116.241103} {\bibfield
  {journal} {\bibinfo  {journal} {Phys. Rev. Lett.}\ }\textbf {\bibinfo
  {volume} {116}},\ \bibinfo {pages} {241103} (\bibinfo {year}
  {2016}{\natexlab{b}})},\ \Eprint {http://arxiv.org/abs/1606.04855}
  {arXiv:1606.04855 [gr-qc]} \BibitemShut {NoStop}%
%%CITATION = ARXIV:1606.04855;%%
\bibitem [{\citenamefont {Abbott}\ \emph
  {et~al.}(2017{\natexlab{a}})\citenamefont {Abbott} \emph
  {et~al.}}]{Abbott:2017vtc}%
  \BibitemOpen
  \bibfield  {author} {\bibinfo {author} {\bibfnamefont {B.~P.}\ \bibnamefont
  {Abbott}} \emph {et~al.} (\bibinfo {collaboration} {VIRGO, LIGO
  Scientific}),\ }\href {\doibase 10.1103/PhysRevLett.118.221101} {\bibfield
  {journal} {\bibinfo  {journal} {Phys. Rev. Lett.}\ }\textbf {\bibinfo
  {volume} {118}},\ \bibinfo {pages} {221101} (\bibinfo {year}
  {2017}{\natexlab{a}})},\ \Eprint {http://arxiv.org/abs/1706.01812}
  {arXiv:1706.01812 [gr-qc]} \BibitemShut {NoStop}%
%%CITATION = ARXIV:1706.01812;%%
\bibitem [{\citenamefont {Abbott}\ \emph
  {et~al.}(2017{\natexlab{b}})\citenamefont {Abbott} \emph
  {et~al.}}]{Abbott:2017oio}%
  \BibitemOpen
  \bibfield  {author} {\bibinfo {author} {\bibfnamefont {B.~P.}\ \bibnamefont
  {Abbott}} \emph {et~al.} (\bibinfo {collaboration} {Virgo, LIGO
  Scientific}),\ }\href@noop {} {\bibfield  {journal} {\bibinfo  {journal}
  {Submitted to: Phys. Rev. Lett.}\ } (\bibinfo {year} {2017}{\natexlab{b}})},\
  \Eprint {http://arxiv.org/abs/1709.09660} {arXiv:1709.09660 [gr-qc]}
  \BibitemShut {NoStop}%
%%CITATION = ARXIV:1709.09660;%%
\bibitem [{\citenamefont {Abbott}\ \emph
  {et~al.}(2016{\natexlab{c}})\citenamefont {Abbott} \emph
  {et~al.}}]{TheLIGOScientific:2016src}%
  \BibitemOpen
  \bibfield  {author} {\bibinfo {author} {\bibfnamefont {B.~P.}\ \bibnamefont
  {Abbott}} \emph {et~al.} (\bibinfo {collaboration} {Virgo, LIGO
  Scientific}),\ }\href {\doibase 10.1103/PhysRevLett.116.221101} {\bibfield
  {journal} {\bibinfo  {journal} {Phys. Rev. Lett.}\ }\textbf {\bibinfo
  {volume} {116}},\ \bibinfo {pages} {221101} (\bibinfo {year}
  {2016}{\natexlab{c}})},\ \Eprint {http://arxiv.org/abs/1602.03841}
  {arXiv:1602.03841 [gr-qc]} \BibitemShut {NoStop}%
%%CITATION = ARXIV:1602.03841;%%
\bibitem [{\citenamefont {Yunes}\ \emph {et~al.}(2016)\citenamefont {Yunes},
  \citenamefont {Yagi},\ and\ \citenamefont {Pretorius}}]{Yunes:2016jcc}%
  \BibitemOpen
  \bibfield  {author} {\bibinfo {author} {\bibfnamefont {N.}~\bibnamefont
  {Yunes}}, \bibinfo {author} {\bibfnamefont {K.}~\bibnamefont {Yagi}}, \ and\
  \bibinfo {author} {\bibfnamefont {F.}~\bibnamefont {Pretorius}},\ }\href
  {\doibase 10.1103/PhysRevD.94.084002} {\bibfield  {journal} {\bibinfo
  {journal} {Phys. Rev.}\ }\textbf {\bibinfo {volume} {D94}},\ \bibinfo {pages}
  {084002} (\bibinfo {year} {2016})},\ \Eprint
  {http://arxiv.org/abs/1603.08955} {arXiv:1603.08955 [gr-qc]} \BibitemShut
  {NoStop}%
%%CITATION = ARXIV:1603.08955;%%
\bibitem [{\citenamefont {Gair}\ \emph {et~al.}(2013)\citenamefont {Gair},
  \citenamefont {Vallisneri}, \citenamefont {Larson},\ and\ \citenamefont
  {Baker}}]{Gair:2012nm}%
  \BibitemOpen
  \bibfield  {author} {\bibinfo {author} {\bibfnamefont {J.~R.}\ \bibnamefont
  {Gair}}, \bibinfo {author} {\bibfnamefont {M.}~\bibnamefont {Vallisneri}},
  \bibinfo {author} {\bibfnamefont {S.~L.}\ \bibnamefont {Larson}}, \ and\
  \bibinfo {author} {\bibfnamefont {J.~G.}\ \bibnamefont {Baker}},\ }\href
  {\doibase 10.12942/lrr-2013-7} {\bibfield  {journal} {\bibinfo  {journal}
  {Living Rev. Rel.}\ }\textbf {\bibinfo {volume} {16}},\ \bibinfo {pages} {7}
  (\bibinfo {year} {2013})},\ \Eprint {http://arxiv.org/abs/1212.5575}
  {arXiv:1212.5575 [gr-qc]} \BibitemShut {NoStop}%
%%CITATION = ARXIV:1212.5575;%%
\bibitem [{\citenamefont {Yunes}\ and\ \citenamefont
  {Siemens}(2013)}]{Yunes:2013dva}%
  \BibitemOpen
  \bibfield  {author} {\bibinfo {author} {\bibfnamefont {N.}~\bibnamefont
  {Yunes}}\ and\ \bibinfo {author} {\bibfnamefont {X.}~\bibnamefont
  {Siemens}},\ }\href {\doibase 10.12942/lrr-2013-9} {\bibfield  {journal}
  {\bibinfo  {journal} {Living Rev. Rel.}\ }\textbf {\bibinfo {volume} {16}},\
  \bibinfo {pages} {9} (\bibinfo {year} {2013})},\ \Eprint
  {http://arxiv.org/abs/1304.3473} {arXiv:1304.3473 [gr-qc]} \BibitemShut
  {NoStop}%
%%CITATION = ARXIV:1304.3473;%%
\bibitem [{\citenamefont {Berti}\ \emph {et~al.}(2015)\citenamefont {Berti}
  \emph {et~al.}}]{Berti:2015itd}%
  \BibitemOpen
  \bibfield  {author} {\bibinfo {author} {\bibfnamefont {E.}~\bibnamefont
  {Berti}} \emph {et~al.},\ }\href {\doibase 10.1088/0264-9381/32/24/243001}
  {\bibfield  {journal} {\bibinfo  {journal} {Class. Quant. Grav.}\ }\textbf
  {\bibinfo {volume} {32}},\ \bibinfo {pages} {243001} (\bibinfo {year}
  {2015})},\ \Eprint {http://arxiv.org/abs/1501.07274} {arXiv:1501.07274
  [gr-qc]} \BibitemShut {NoStop}%
%%CITATION = ARXIV:1501.07274;%%
\bibitem [{\citenamefont {Detweiler}(1980)}]{Detweiler:1980gk}%
  \BibitemOpen
  \bibfield  {author} {\bibinfo {author} {\bibfnamefont {S.~L.}\ \bibnamefont
  {Detweiler}},\ }\href {\doibase 10.1086/158109} {\bibfield  {journal}
  {\bibinfo  {journal} {Astrophys. J.}\ }\textbf {\bibinfo {volume} {239}},\
  \bibinfo {pages} {292} (\bibinfo {year} {1980})}\BibitemShut {NoStop}%
%%CITATION = ASJOA,239,292;%%
\bibitem [{\citenamefont {Dreyer}\ \emph {et~al.}(2004)\citenamefont {Dreyer},
  \citenamefont {Kelly}, \citenamefont {Krishnan}, \citenamefont {Finn},
  \citenamefont {Garrison},\ and\ \citenamefont
  {Lopez-Aleman}}]{Dreyer:2003bv}%
  \BibitemOpen
  \bibfield  {author} {\bibinfo {author} {\bibfnamefont {O.}~\bibnamefont
  {Dreyer}}, \bibinfo {author} {\bibfnamefont {B.~J.}\ \bibnamefont {Kelly}},
  \bibinfo {author} {\bibfnamefont {B.}~\bibnamefont {Krishnan}}, \bibinfo
  {author} {\bibfnamefont {L.~S.}\ \bibnamefont {Finn}}, \bibinfo {author}
  {\bibfnamefont {D.}~\bibnamefont {Garrison}}, \ and\ \bibinfo {author}
  {\bibfnamefont {R.}~\bibnamefont {Lopez-Aleman}},\ }\href {\doibase
  10.1088/0264-9381/21/4/003} {\bibfield  {journal} {\bibinfo  {journal}
  {Class. Quant. Grav.}\ }\textbf {\bibinfo {volume} {21}},\ \bibinfo {pages}
  {787} (\bibinfo {year} {2004})},\ \Eprint
  {http://arxiv.org/abs/gr-qc/0309007} {arXiv:gr-qc/0309007 [gr-qc]}
  \BibitemShut {NoStop}%
%%CITATION = GR-QC/0309007;%%
\bibitem [{\citenamefont {Berti}\ \emph
  {et~al.}(2006{\natexlab{a}})\citenamefont {Berti}, \citenamefont {Cardoso},\
  and\ \citenamefont {Will}}]{Berti:2005ys}%
  \BibitemOpen
  \bibfield  {author} {\bibinfo {author} {\bibfnamefont {E.}~\bibnamefont
  {Berti}}, \bibinfo {author} {\bibfnamefont {V.}~\bibnamefont {Cardoso}}, \
  and\ \bibinfo {author} {\bibfnamefont {C.~M.}\ \bibnamefont {Will}},\ }\href
  {\doibase 10.1103/PhysRevD.73.064030} {\bibfield  {journal} {\bibinfo
  {journal} {Phys. Rev.}\ }\textbf {\bibinfo {volume} {D73}},\ \bibinfo {pages}
  {064030} (\bibinfo {year} {2006}{\natexlab{a}})},\ \Eprint
  {http://arxiv.org/abs/gr-qc/0512160} {arXiv:gr-qc/0512160 [gr-qc]}
  \BibitemShut {NoStop}%
%%CITATION = GR-QC/0512160;%%
\bibitem [{\citenamefont {Berti}\ \emph {et~al.}(2009)\citenamefont {Berti},
  \citenamefont {Cardoso},\ and\ \citenamefont {Starinets}}]{Berti:2009kk}%
  \BibitemOpen
  \bibfield  {author} {\bibinfo {author} {\bibfnamefont {E.}~\bibnamefont
  {Berti}}, \bibinfo {author} {\bibfnamefont {V.}~\bibnamefont {Cardoso}}, \
  and\ \bibinfo {author} {\bibfnamefont {A.~O.}\ \bibnamefont {Starinets}},\
  }\href {\doibase 10.1088/0264-9381/26/16/163001} {\bibfield  {journal}
  {\bibinfo  {journal} {Class. Quant. Grav.}\ }\textbf {\bibinfo {volume}
  {26}},\ \bibinfo {pages} {163001} (\bibinfo {year} {2009})},\ \Eprint
  {http://arxiv.org/abs/0905.2975} {arXiv:0905.2975 [gr-qc]} \BibitemShut
  {NoStop}%
%%CITATION = ARXIV:0905.2975;%%
\bibitem [{\citenamefont {Blázquez-Salcedo}\ \emph {et~al.}(2016)\citenamefont
  {Blázquez-Salcedo}, \citenamefont {Macedo}, \citenamefont {Cardoso},
  \citenamefont {Ferrari}, \citenamefont {Gualtieri}, \citenamefont {Khoo},
  \citenamefont {Kunz},\ and\ \citenamefont {Pani}}]{Blazquez-Salcedo:2016enn}%
  \BibitemOpen
  \bibfield  {author} {\bibinfo {author} {\bibfnamefont {J.~L.}\ \bibnamefont
  {Blázquez-Salcedo}}, \bibinfo {author} {\bibfnamefont {C.~F.~B.}\
  \bibnamefont {Macedo}}, \bibinfo {author} {\bibfnamefont {V.}~\bibnamefont
  {Cardoso}}, \bibinfo {author} {\bibfnamefont {V.}~\bibnamefont {Ferrari}},
  \bibinfo {author} {\bibfnamefont {L.}~\bibnamefont {Gualtieri}}, \bibinfo
  {author} {\bibfnamefont {F.~S.}\ \bibnamefont {Khoo}}, \bibinfo {author}
  {\bibfnamefont {J.}~\bibnamefont {Kunz}}, \ and\ \bibinfo {author}
  {\bibfnamefont {P.}~\bibnamefont {Pani}},\ }\href {\doibase
  10.1103/PhysRevD.94.104024} {\bibfield  {journal} {\bibinfo  {journal} {Phys.
  Rev.}\ }\textbf {\bibinfo {volume} {D94}},\ \bibinfo {pages} {104024}
  (\bibinfo {year} {2016})},\ \Eprint {http://arxiv.org/abs/1609.01286}
  {arXiv:1609.01286 [gr-qc]} \BibitemShut {NoStop}%
%%CITATION = ARXIV:1609.01286;%%
\bibitem [{\citenamefont {Glampedakis}\ \emph {et~al.}(2017)\citenamefont
  {Glampedakis}, \citenamefont {Pappas}, \citenamefont {Silva},\ and\
  \citenamefont {Berti}}]{Glampedakis:2017dvb}%
  \BibitemOpen
  \bibfield  {author} {\bibinfo {author} {\bibfnamefont {K.}~\bibnamefont
  {Glampedakis}}, \bibinfo {author} {\bibfnamefont {G.}~\bibnamefont {Pappas}},
  \bibinfo {author} {\bibfnamefont {H.~O.}\ \bibnamefont {Silva}}, \ and\
  \bibinfo {author} {\bibfnamefont {E.}~\bibnamefont {Berti}},\ }\href@noop {}
  {\  (\bibinfo {year} {2017})},\ \Eprint {http://arxiv.org/abs/1706.07658}
  {arXiv:1706.07658 [gr-qc]} \BibitemShut {NoStop}%
%%CITATION = ARXIV:1706.07658;%%
\bibitem [{\citenamefont {Cardoso}\ and\ \citenamefont
  {Gualtieri}(2016)}]{Cardoso:2016ryw}%
  \BibitemOpen
  \bibfield  {author} {\bibinfo {author} {\bibfnamefont {V.}~\bibnamefont
  {Cardoso}}\ and\ \bibinfo {author} {\bibfnamefont {L.}~\bibnamefont
  {Gualtieri}},\ }\href {\doibase 10.1088/0264-9381/33/17/174001} {\bibfield
  {journal} {\bibinfo  {journal} {Class. Quant. Grav.}\ }\textbf {\bibinfo
  {volume} {33}},\ \bibinfo {pages} {174001} (\bibinfo {year} {2016})},\
  \Eprint {http://arxiv.org/abs/1607.03133} {arXiv:1607.03133 [gr-qc]}
  \BibitemShut {NoStop}%
%%CITATION = ARXIV:1607.03133;%%
\bibitem [{\citenamefont {Cardoso}\ and\ \citenamefont
  {Pani}(2017)}]{Cardoso:2017njb}%
  \BibitemOpen
  \bibfield  {author} {\bibinfo {author} {\bibfnamefont {V.}~\bibnamefont
  {Cardoso}}\ and\ \bibinfo {author} {\bibfnamefont {P.}~\bibnamefont {Pani}},\
  }\href@noop {} {\  (\bibinfo {year} {2017})},\ \Eprint
  {http://arxiv.org/abs/1707.03021} {arXiv:1707.03021 [gr-qc]} \BibitemShut
  {NoStop}%
%%CITATION = ARXIV:1707.03021;%%
\bibitem [{\citenamefont {Cardoso}\ \emph
  {et~al.}(2016{\natexlab{a}})\citenamefont {Cardoso}, \citenamefont
  {Franzin},\ and\ \citenamefont {Pani}}]{Cardoso:2016rao}%
  \BibitemOpen
  \bibfield  {author} {\bibinfo {author} {\bibfnamefont {V.}~\bibnamefont
  {Cardoso}}, \bibinfo {author} {\bibfnamefont {E.}~\bibnamefont {Franzin}}, \
  and\ \bibinfo {author} {\bibfnamefont {P.}~\bibnamefont {Pani}},\ }\href
  {\doibase 10.1103/PhysRevLett.117.089902, 10.1103/PhysRevLett.116.171101}
  {\bibfield  {journal} {\bibinfo  {journal} {Phys. Rev. Lett.}\ }\textbf
  {\bibinfo {volume} {116}},\ \bibinfo {pages} {171101} (\bibinfo {year}
  {2016}{\natexlab{a}})},\ \bibinfo {note} {[Erratum: Phys. Rev.
  Lett.117,no.8,089902(2016)]},\ \Eprint {http://arxiv.org/abs/1602.07309}
  {arXiv:1602.07309 [gr-qc]} \BibitemShut {NoStop}%
%%CITATION = ARXIV:1602.07309;%%
\bibitem [{\citenamefont {Cardoso}\ \emph
  {et~al.}(2016{\natexlab{b}})\citenamefont {Cardoso}, \citenamefont {Hopper},
  \citenamefont {Macedo}, \citenamefont {Palenzuela},\ and\ \citenamefont
  {Pani}}]{Cardoso:2016oxy}%
  \BibitemOpen
  \bibfield  {author} {\bibinfo {author} {\bibfnamefont {V.}~\bibnamefont
  {Cardoso}}, \bibinfo {author} {\bibfnamefont {S.}~\bibnamefont {Hopper}},
  \bibinfo {author} {\bibfnamefont {C.~F.~B.}\ \bibnamefont {Macedo}}, \bibinfo
  {author} {\bibfnamefont {C.}~\bibnamefont {Palenzuela}}, \ and\ \bibinfo
  {author} {\bibfnamefont {P.}~\bibnamefont {Pani}},\ }\href {\doibase
  10.1103/PhysRevD.94.084031} {\bibfield  {journal} {\bibinfo  {journal} {Phys.
  Rev.}\ }\textbf {\bibinfo {volume} {D94}},\ \bibinfo {pages} {084031}
  (\bibinfo {year} {2016}{\natexlab{b}})},\ \Eprint
  {http://arxiv.org/abs/1608.08637} {arXiv:1608.08637 [gr-qc]} \BibitemShut
  {NoStop}%
%%CITATION = ARXIV:1608.08637;%%
\bibitem [{\citenamefont {Abedi}\ \emph {et~al.}(2016)\citenamefont {Abedi},
  \citenamefont {Dykaar},\ and\ \citenamefont {Afshordi}}]{Abedi:2016hgu}%
  \BibitemOpen
  \bibfield  {author} {\bibinfo {author} {\bibfnamefont {J.}~\bibnamefont
  {Abedi}}, \bibinfo {author} {\bibfnamefont {H.}~\bibnamefont {Dykaar}}, \
  and\ \bibinfo {author} {\bibfnamefont {N.}~\bibnamefont {Afshordi}},\
  }\href@noop {} {\  (\bibinfo {year} {2016})},\ \Eprint
  {http://arxiv.org/abs/1612.00266} {arXiv:1612.00266 [gr-qc]} \BibitemShut
  {NoStop}%
%%CITATION = ARXIV:1612.00266;%%
\bibitem [{\citenamefont {Mark}\ \emph {et~al.}(2017)\citenamefont {Mark},
  \citenamefont {Zimmerman}, \citenamefont {Du},\ and\ \citenamefont
  {Chen}}]{Mark:2017dnq}%
  \BibitemOpen
  \bibfield  {author} {\bibinfo {author} {\bibfnamefont {Z.}~\bibnamefont
  {Mark}}, \bibinfo {author} {\bibfnamefont {A.}~\bibnamefont {Zimmerman}},
  \bibinfo {author} {\bibfnamefont {S.~M.}\ \bibnamefont {Du}}, \ and\ \bibinfo
  {author} {\bibfnamefont {Y.}~\bibnamefont {Chen}},\ }\href@noop {} {\
  (\bibinfo {year} {2017})},\ \Eprint {http://arxiv.org/abs/1706.06155}
  {arXiv:1706.06155 [gr-qc]} \BibitemShut {NoStop}%
%%CITATION = ARXIV:1706.06155;%%
\bibitem [{\citenamefont {Berti}\ \emph
  {et~al.}(2007{\natexlab{a}})\citenamefont {Berti}, \citenamefont {Cardoso},
  \citenamefont {Cardoso},\ and\ \citenamefont {Cavaglia}}]{Berti:2007zu}%
  \BibitemOpen
  \bibfield  {author} {\bibinfo {author} {\bibfnamefont {E.}~\bibnamefont
  {Berti}}, \bibinfo {author} {\bibfnamefont {J.}~\bibnamefont {Cardoso}},
  \bibinfo {author} {\bibfnamefont {V.}~\bibnamefont {Cardoso}}, \ and\
  \bibinfo {author} {\bibfnamefont {M.}~\bibnamefont {Cavaglia}},\ }\href
  {\doibase 10.1103/PhysRevD.76.104044} {\bibfield  {journal} {\bibinfo
  {journal} {Phys. Rev.}\ }\textbf {\bibinfo {volume} {D76}},\ \bibinfo {pages}
  {104044} (\bibinfo {year} {2007}{\natexlab{a}})},\ \Eprint
  {http://arxiv.org/abs/0707.1202} {arXiv:0707.1202 [gr-qc]} \BibitemShut
  {NoStop}%
%%CITATION = ARXIV:0707.1202;%%
\bibitem [{\citenamefont {Berti}\ \emph {et~al.}(2016)\citenamefont {Berti},
  \citenamefont {Sesana}, \citenamefont {Barausse}, \citenamefont {Cardoso},\
  and\ \citenamefont {Belczynski}}]{Berti:2016lat}%
  \BibitemOpen
  \bibfield  {author} {\bibinfo {author} {\bibfnamefont {E.}~\bibnamefont
  {Berti}}, \bibinfo {author} {\bibfnamefont {A.}~\bibnamefont {Sesana}},
  \bibinfo {author} {\bibfnamefont {E.}~\bibnamefont {Barausse}}, \bibinfo
  {author} {\bibfnamefont {V.}~\bibnamefont {Cardoso}}, \ and\ \bibinfo
  {author} {\bibfnamefont {K.}~\bibnamefont {Belczynski}},\ }\href {\doibase
  10.1103/PhysRevLett.117.101102} {\bibfield  {journal} {\bibinfo  {journal}
  {Phys. Rev. Lett.}\ }\textbf {\bibinfo {volume} {117}},\ \bibinfo {pages}
  {101102} (\bibinfo {year} {2016})},\ \Eprint
  {http://arxiv.org/abs/1605.09286} {arXiv:1605.09286 [gr-qc]} \BibitemShut
  {NoStop}%
%%CITATION = ARXIV:1605.09286;%%
\bibitem [{\citenamefont {Bhagwat}\ \emph {et~al.}(2016)\citenamefont
  {Bhagwat}, \citenamefont {Brown},\ and\ \citenamefont
  {Ballmer}}]{Bhagwat:2016ntk}%
  \BibitemOpen
  \bibfield  {author} {\bibinfo {author} {\bibfnamefont {S.}~\bibnamefont
  {Bhagwat}}, \bibinfo {author} {\bibfnamefont {D.~A.}\ \bibnamefont {Brown}},
  \ and\ \bibinfo {author} {\bibfnamefont {S.~W.}\ \bibnamefont {Ballmer}},\
  }\href {\doibase 10.1103/PhysRevD.94.084024, 10.1103/PhysRevD.95.069906}
  {\bibfield  {journal} {\bibinfo  {journal} {Phys. Rev.}\ }\textbf {\bibinfo
  {volume} {D94}},\ \bibinfo {pages} {084024} (\bibinfo {year} {2016})},\
  \bibinfo {note} {[Erratum: Phys. Rev.D95,no.6,069906(2017)]},\ \Eprint
  {http://arxiv.org/abs/1607.07845} {arXiv:1607.07845 [gr-qc]} \BibitemShut
  {NoStop}%
%%CITATION = ARXIV:1607.07845;%%
\bibitem [{\citenamefont {Price}(1972)}]{Price:1971fb}%
  \BibitemOpen
  \bibfield  {author} {\bibinfo {author} {\bibfnamefont {R.~H.}\ \bibnamefont
  {Price}},\ }\href {\doibase 10.1103/PhysRevD.5.2419} {\bibfield  {journal}
  {\bibinfo  {journal} {Phys. Rev.}\ }\textbf {\bibinfo {volume} {D5}},\
  \bibinfo {pages} {2419} (\bibinfo {year} {1972})}\BibitemShut {NoStop}%
%%CITATION = PHRVA,D5,2419;%%
\bibitem [{\citenamefont {Thrane}\ \emph {et~al.}(2017)\citenamefont {Thrane},
  \citenamefont {Lasky},\ and\ \citenamefont {Levin}}]{Thrane:2017lqn}%
  \BibitemOpen
  \bibfield  {author} {\bibinfo {author} {\bibfnamefont {E.}~\bibnamefont
  {Thrane}}, \bibinfo {author} {\bibfnamefont {P.~D.}\ \bibnamefont {Lasky}}, \
  and\ \bibinfo {author} {\bibfnamefont {Y.}~\bibnamefont {Levin}},\
  }\href@noop {} {\  (\bibinfo {year} {2017})},\ \Eprint
  {http://arxiv.org/abs/1706.05152} {arXiv:1706.05152 [gr-qc]} \BibitemShut
  {NoStop}%
%%CITATION = ARXIV:1706.05152;%%
\bibitem [{\citenamefont {Buonanno}\ \emph {et~al.}(2009)\citenamefont
  {Buonanno}, \citenamefont {Pan}, \citenamefont {Pfeiffer}, \citenamefont
  {Scheel}, \citenamefont {Buchman},\ and\ \citenamefont
  {Kidder}}]{Buonanno:2009qa}%
  \BibitemOpen
  \bibfield  {author} {\bibinfo {author} {\bibfnamefont {A.}~\bibnamefont
  {Buonanno}}, \bibinfo {author} {\bibfnamefont {Y.}~\bibnamefont {Pan}},
  \bibinfo {author} {\bibfnamefont {H.~P.}\ \bibnamefont {Pfeiffer}}, \bibinfo
  {author} {\bibfnamefont {M.~A.}\ \bibnamefont {Scheel}}, \bibinfo {author}
  {\bibfnamefont {L.~T.}\ \bibnamefont {Buchman}}, \ and\ \bibinfo {author}
  {\bibfnamefont {L.~E.}\ \bibnamefont {Kidder}},\ }\href {\doibase
  10.1103/PhysRevD.79.124028} {\bibfield  {journal} {\bibinfo  {journal} {Phys.
  Rev.}\ }\textbf {\bibinfo {volume} {D79}},\ \bibinfo {pages} {124028}
  (\bibinfo {year} {2009})},\ \Eprint {http://arxiv.org/abs/0902.0790}
  {arXiv:0902.0790 [gr-qc]} \BibitemShut {NoStop}%
%%CITATION = ARXIV:0902.0790;%%
\bibitem [{\citenamefont {Zimmerman}\ and\ \citenamefont
  {Chen}(2011)}]{Zimmerman:2011dx}%
  \BibitemOpen
  \bibfield  {author} {\bibinfo {author} {\bibfnamefont {A.}~\bibnamefont
  {Zimmerman}}\ and\ \bibinfo {author} {\bibfnamefont {Y.}~\bibnamefont
  {Chen}},\ }\href {\doibase 10.1103/PhysRevD.84.084012} {\bibfield  {journal}
  {\bibinfo  {journal} {Phys. Rev.}\ }\textbf {\bibinfo {volume} {D84}},\
  \bibinfo {pages} {084012} (\bibinfo {year} {2011})},\ \Eprint
  {http://arxiv.org/abs/1106.0782} {arXiv:1106.0782 [gr-qc]} \BibitemShut
  {NoStop}%
%%CITATION = ARXIV:1106.0782;%%
\bibitem [{\citenamefont {Mroue}\ \emph {et~al.}(2013)\citenamefont {Mroue}
  \emph {et~al.}}]{Mroue:2013xna}%
  \BibitemOpen
  \bibfield  {author} {\bibinfo {author} {\bibfnamefont {A.~H.}\ \bibnamefont
  {Mroue}} \emph {et~al.},\ }\href {\doibase 10.1103/PhysRevLett.111.241104}
  {\bibfield  {journal} {\bibinfo  {journal} {Phys. Rev. Lett.}\ }\textbf
  {\bibinfo {volume} {111}},\ \bibinfo {pages} {241104} (\bibinfo {year}
  {2013})},\ \Eprint {http://arxiv.org/abs/1304.6077} {arXiv:1304.6077 [gr-qc]}
  \BibitemShut {NoStop}%
%%CITATION = ARXIV:1304.6077;%%
\bibitem [{\citenamefont {Sundararajan}\ \emph {et~al.}(2010)\citenamefont
  {Sundararajan}, \citenamefont {Khanna},\ and\ \citenamefont
  {Hughes}}]{Sundararajan:2010sr}%
  \BibitemOpen
  \bibfield  {author} {\bibinfo {author} {\bibfnamefont {P.~A.}\ \bibnamefont
  {Sundararajan}}, \bibinfo {author} {\bibfnamefont {G.}~\bibnamefont
  {Khanna}}, \ and\ \bibinfo {author} {\bibfnamefont {S.~A.}\ \bibnamefont
  {Hughes}},\ }\href {\doibase 10.1103/PhysRevD.81.104009} {\bibfield
  {journal} {\bibinfo  {journal} {Phys. Rev.}\ }\textbf {\bibinfo {volume}
  {D81}},\ \bibinfo {pages} {104009} (\bibinfo {year} {2010})},\ \Eprint
  {http://arxiv.org/abs/1003.0485} {arXiv:1003.0485 [gr-qc]} \BibitemShut
  {NoStop}%
%%CITATION = ARXIV:1003.0485;%%
\bibitem [{\citenamefont {Zenginoglu}\ and\ \citenamefont
  {Khanna}(2011)}]{Zenginoglu:2011zz}%
  \BibitemOpen
  \bibfield  {author} {\bibinfo {author} {\bibfnamefont {A.}~\bibnamefont
  {Zenginoglu}}\ and\ \bibinfo {author} {\bibfnamefont {G.}~\bibnamefont
  {Khanna}},\ }\href {\doibase 10.1103/PhysRevX.1.021017} {\bibfield  {journal}
  {\bibinfo  {journal} {Phys. Rev.}\ }\textbf {\bibinfo {volume} {X1}},\
  \bibinfo {pages} {021017} (\bibinfo {year} {2011})},\ \Eprint
  {http://arxiv.org/abs/1108.1816} {arXiv:1108.1816 [gr-qc]} \BibitemShut
  {NoStop}%
%%CITATION = ARXIV:1108.1816;%%
\bibitem [{\citenamefont {Meidam}\ \emph {et~al.}(2014)\citenamefont {Meidam},
  \citenamefont {Agathos}, \citenamefont {Van Den~Broeck}, \citenamefont
  {Veitch},\ and\ \citenamefont {Sathyaprakash}}]{Meidam:2014jpa}%
  \BibitemOpen
  \bibfield  {author} {\bibinfo {author} {\bibfnamefont {J.}~\bibnamefont
  {Meidam}}, \bibinfo {author} {\bibfnamefont {M.}~\bibnamefont {Agathos}},
  \bibinfo {author} {\bibfnamefont {C.}~\bibnamefont {Van Den~Broeck}},
  \bibinfo {author} {\bibfnamefont {J.}~\bibnamefont {Veitch}}, \ and\ \bibinfo
  {author} {\bibfnamefont {B.~S.}\ \bibnamefont {Sathyaprakash}},\ }\href
  {\doibase 10.1103/PhysRevD.90.064009} {\bibfield  {journal} {\bibinfo
  {journal} {Phys. Rev.}\ }\textbf {\bibinfo {volume} {D90}},\ \bibinfo {pages}
  {064009} (\bibinfo {year} {2014})},\ \Eprint {http://arxiv.org/abs/1406.3201}
  {arXiv:1406.3201 [gr-qc]} \BibitemShut {NoStop}%
%%CITATION = ARXIV:1406.3201;%%
\bibitem [{\citenamefont {Yang}\ \emph {et~al.}(2017)\citenamefont {Yang},
  \citenamefont {Yagi}, \citenamefont {Blackman}, \citenamefont {Lehner},
  \citenamefont {Paschalidis}, \citenamefont {Pretorius},\ and\ \citenamefont
  {Yunes}}]{Yang:2017zxs}%
  \BibitemOpen
  \bibfield  {author} {\bibinfo {author} {\bibfnamefont {H.}~\bibnamefont
  {Yang}}, \bibinfo {author} {\bibfnamefont {K.}~\bibnamefont {Yagi}}, \bibinfo
  {author} {\bibfnamefont {J.}~\bibnamefont {Blackman}}, \bibinfo {author}
  {\bibfnamefont {L.}~\bibnamefont {Lehner}}, \bibinfo {author} {\bibfnamefont
  {V.}~\bibnamefont {Paschalidis}}, \bibinfo {author} {\bibfnamefont
  {F.}~\bibnamefont {Pretorius}}, \ and\ \bibinfo {author} {\bibfnamefont
  {N.}~\bibnamefont {Yunes}},\ }\href {\doibase 10.1103/PhysRevLett.118.161101}
  {\bibfield  {journal} {\bibinfo  {journal} {Phys. Rev. Lett.}\ }\textbf
  {\bibinfo {volume} {118}},\ \bibinfo {pages} {161101} (\bibinfo {year}
  {2017})},\ \Eprint {http://arxiv.org/abs/1701.05808} {arXiv:1701.05808
  [gr-qc]} \BibitemShut {NoStop}%
%%CITATION = ARXIV:1701.05808;%%
\bibitem [{\citenamefont {Andersson}(1995)}]{Andersson:1995zk}%
  \BibitemOpen
  \bibfield  {author} {\bibinfo {author} {\bibfnamefont {N.}~\bibnamefont
  {Andersson}},\ }\href {\doibase 10.1103/PhysRevD.51.353} {\bibfield
  {journal} {\bibinfo  {journal} {Phys. Rev.}\ }\textbf {\bibinfo {volume}
  {D51}},\ \bibinfo {pages} {353} (\bibinfo {year} {1995})}\BibitemShut
  {NoStop}%
%%CITATION = PHRVA,D51,353;%%
\bibitem [{\citenamefont {Nollert}\ and\ \citenamefont
  {Price}(1999)}]{Nollert:1998ys}%
  \BibitemOpen
  \bibfield  {author} {\bibinfo {author} {\bibfnamefont {H.-P.}\ \bibnamefont
  {Nollert}}\ and\ \bibinfo {author} {\bibfnamefont {R.~H.}\ \bibnamefont
  {Price}},\ }\href {\doibase 10.1063/1.532698} {\bibfield  {journal} {\bibinfo
   {journal} {J. Math. Phys.}\ }\textbf {\bibinfo {volume} {40}},\ \bibinfo
  {pages} {980} (\bibinfo {year} {1999})},\ \Eprint
  {http://arxiv.org/abs/gr-qc/9810074} {arXiv:gr-qc/9810074 [gr-qc]}
  \BibitemShut {NoStop}%
%%CITATION = GR-QC/9810074;%%
\bibitem [{\citenamefont {Nollert}(2000{\natexlab{a}})}]{nollertthesis}%
  \BibitemOpen
  \bibfield  {author} {\bibinfo {author} {\bibfnamefont {H.-P.}\ \bibnamefont
  {Nollert}},\ }\href@noop {} {\emph {\bibinfo {title} {Characteristic
  Oscillations of Black Holes and Neutron Stars: From Mathematical Background
  to Astrophysical Applications}}}\ (\bibinfo  {publisher}
  {Habilitationsschrift Der Fakultat fur Physik der Eberhard-Karls-Universitat,
  Tubingen},\ \bibinfo {address} {Tubingen},\ \bibinfo {year}
  {2000})\BibitemShut {NoStop}%
\bibitem [{\citenamefont {Berti}\ and\ \citenamefont
  {Cardoso}(2006)}]{Berti:2006wq}%
  \BibitemOpen
  \bibfield  {author} {\bibinfo {author} {\bibfnamefont {E.}~\bibnamefont
  {Berti}}\ and\ \bibinfo {author} {\bibfnamefont {V.}~\bibnamefont
  {Cardoso}},\ }\href {\doibase 10.1103/PhysRevD.74.104020} {\bibfield
  {journal} {\bibinfo  {journal} {Phys. Rev.}\ }\textbf {\bibinfo {volume}
  {D74}},\ \bibinfo {pages} {104020} (\bibinfo {year} {2006})},\ \Eprint
  {http://arxiv.org/abs/gr-qc/0605118} {arXiv:gr-qc/0605118 [gr-qc]}
  \BibitemShut {NoStop}%
%%CITATION = GR-QC/0605118;%%
\bibitem [{\citenamefont {Zhang}\ \emph {et~al.}(2013)\citenamefont {Zhang},
  \citenamefont {Berti},\ and\ \citenamefont {Cardoso}}]{Zhang:2013ksa}%
  \BibitemOpen
  \bibfield  {author} {\bibinfo {author} {\bibfnamefont {Z.}~\bibnamefont
  {Zhang}}, \bibinfo {author} {\bibfnamefont {E.}~\bibnamefont {Berti}}, \ and\
  \bibinfo {author} {\bibfnamefont {V.}~\bibnamefont {Cardoso}},\ }\href
  {\doibase 10.1103/PhysRevD.88.044018} {\bibfield  {journal} {\bibinfo
  {journal} {Phys. Rev.}\ }\textbf {\bibinfo {volume} {D88}},\ \bibinfo {pages}
  {044018} (\bibinfo {year} {2013})},\ \Eprint {http://arxiv.org/abs/1305.4306}
  {arXiv:1305.4306 [gr-qc]} \BibitemShut {NoStop}%
%%CITATION = ARXIV:1305.4306;%%
\bibitem [{\citenamefont {Krivan}\ \emph {et~al.}(1997)\citenamefont {Krivan},
  \citenamefont {Laguna}, \citenamefont {Papadopoulos},\ and\ \citenamefont
  {Andersson}}]{Krivan:1997hc}%
  \BibitemOpen
  \bibfield  {author} {\bibinfo {author} {\bibfnamefont {W.}~\bibnamefont
  {Krivan}}, \bibinfo {author} {\bibfnamefont {P.}~\bibnamefont {Laguna}},
  \bibinfo {author} {\bibfnamefont {P.}~\bibnamefont {Papadopoulos}}, \ and\
  \bibinfo {author} {\bibfnamefont {N.}~\bibnamefont {Andersson}},\ }\href
  {\doibase 10.1103/PhysRevD.56.3395} {\bibfield  {journal} {\bibinfo
  {journal} {Phys. Rev.}\ }\textbf {\bibinfo {volume} {D56}},\ \bibinfo {pages}
  {3395} (\bibinfo {year} {1997})},\ \Eprint
  {http://arxiv.org/abs/gr-qc/9702048} {arXiv:gr-qc/9702048 [gr-qc]}
  \BibitemShut {NoStop}%
%%CITATION = GR-QC/9702048;%%
\bibitem [{\citenamefont {Dorband}\ \emph {et~al.}(2006)\citenamefont
  {Dorband}, \citenamefont {Berti}, \citenamefont {Diener}, \citenamefont
  {Schnetter},\ and\ \citenamefont {Tiglio}}]{Dorband:2006gg}%
  \BibitemOpen
  \bibfield  {author} {\bibinfo {author} {\bibfnamefont {E.~N.}\ \bibnamefont
  {Dorband}}, \bibinfo {author} {\bibfnamefont {E.}~\bibnamefont {Berti}},
  \bibinfo {author} {\bibfnamefont {P.}~\bibnamefont {Diener}}, \bibinfo
  {author} {\bibfnamefont {E.}~\bibnamefont {Schnetter}}, \ and\ \bibinfo
  {author} {\bibfnamefont {M.}~\bibnamefont {Tiglio}},\ }\href {\doibase
  10.1103/PhysRevD.74.084028} {\bibfield  {journal} {\bibinfo  {journal} {Phys.
  Rev.}\ }\textbf {\bibinfo {volume} {D74}},\ \bibinfo {pages} {084028}
  (\bibinfo {year} {2006})},\ \Eprint {http://arxiv.org/abs/gr-qc/0608091}
  {arXiv:gr-qc/0608091 [gr-qc]} \BibitemShut {NoStop}%
%%CITATION = GR-QC/0608091;%%
\bibitem [{\citenamefont {Buonanno}\ \emph {et~al.}(2007)\citenamefont
  {Buonanno}, \citenamefont {Cook},\ and\ \citenamefont
  {Pretorius}}]{Buonanno:2006ui}%
  \BibitemOpen
  \bibfield  {author} {\bibinfo {author} {\bibfnamefont {A.}~\bibnamefont
  {Buonanno}}, \bibinfo {author} {\bibfnamefont {G.~B.}\ \bibnamefont {Cook}},
  \ and\ \bibinfo {author} {\bibfnamefont {F.}~\bibnamefont {Pretorius}},\
  }\href {\doibase 10.1103/PhysRevD.75.124018} {\bibfield  {journal} {\bibinfo
  {journal} {Phys. Rev.}\ }\textbf {\bibinfo {volume} {D75}},\ \bibinfo {pages}
  {124018} (\bibinfo {year} {2007})},\ \Eprint
  {http://arxiv.org/abs/gr-qc/0610122} {arXiv:gr-qc/0610122 [gr-qc]}
  \BibitemShut {NoStop}%
%%CITATION = GR-QC/0610122;%%
\bibitem [{\citenamefont {Berti}\ \emph
  {et~al.}(2007{\natexlab{b}})\citenamefont {Berti}, \citenamefont {Cardoso},
  \citenamefont {Gonzalez}, \citenamefont {Sperhake}, \citenamefont {Hannam},
  \citenamefont {Husa},\ and\ \citenamefont {Bruegmann}}]{Berti:2007fi}%
  \BibitemOpen
  \bibfield  {author} {\bibinfo {author} {\bibfnamefont {E.}~\bibnamefont
  {Berti}}, \bibinfo {author} {\bibfnamefont {V.}~\bibnamefont {Cardoso}},
  \bibinfo {author} {\bibfnamefont {J.~A.}\ \bibnamefont {Gonzalez}}, \bibinfo
  {author} {\bibfnamefont {U.}~\bibnamefont {Sperhake}}, \bibinfo {author}
  {\bibfnamefont {M.}~\bibnamefont {Hannam}}, \bibinfo {author} {\bibfnamefont
  {S.}~\bibnamefont {Husa}}, \ and\ \bibinfo {author} {\bibfnamefont
  {B.}~\bibnamefont {Bruegmann}},\ }\href {\doibase 10.1103/PhysRevD.76.064034}
  {\bibfield  {journal} {\bibinfo  {journal} {Phys. Rev.}\ }\textbf {\bibinfo
  {volume} {D76}},\ \bibinfo {pages} {064034} (\bibinfo {year}
  {2007}{\natexlab{b}})},\ \Eprint {http://arxiv.org/abs/gr-qc/0703053}
  {arXiv:gr-qc/0703053 [GR-QC]} \BibitemShut {NoStop}%
%%CITATION = GR-QC/0703053;%%
\bibitem [{\citenamefont {Kamaretsos}\ \emph
  {et~al.}(2012{\natexlab{a}})\citenamefont {Kamaretsos}, \citenamefont
  {Hannam}, \citenamefont {Husa},\ and\ \citenamefont
  {Sathyaprakash}}]{Kamaretsos:2011um}%
  \BibitemOpen
  \bibfield  {author} {\bibinfo {author} {\bibfnamefont {I.}~\bibnamefont
  {Kamaretsos}}, \bibinfo {author} {\bibfnamefont {M.}~\bibnamefont {Hannam}},
  \bibinfo {author} {\bibfnamefont {S.}~\bibnamefont {Husa}}, \ and\ \bibinfo
  {author} {\bibfnamefont {B.~S.}\ \bibnamefont {Sathyaprakash}},\ }\href
  {\doibase 10.1103/PhysRevD.85.024018} {\bibfield  {journal} {\bibinfo
  {journal} {Phys. Rev.}\ }\textbf {\bibinfo {volume} {D85}},\ \bibinfo {pages}
  {024018} (\bibinfo {year} {2012}{\natexlab{a}})},\ \Eprint
  {http://arxiv.org/abs/1107.0854} {arXiv:1107.0854 [gr-qc]} \BibitemShut
  {NoStop}%
%%CITATION = ARXIV:1107.0854;%%
\bibitem [{\citenamefont {London}\ \emph {et~al.}(2014)\citenamefont {London},
  \citenamefont {Shoemaker},\ and\ \citenamefont {Healy}}]{London:2014cma}%
  \BibitemOpen
  \bibfield  {author} {\bibinfo {author} {\bibfnamefont {L.}~\bibnamefont
  {London}}, \bibinfo {author} {\bibfnamefont {D.}~\bibnamefont {Shoemaker}}, \
  and\ \bibinfo {author} {\bibfnamefont {J.}~\bibnamefont {Healy}},\ }\href
  {\doibase 10.1103/PhysRevD.90.124032, 10.1103/PhysRevD.94.069902} {\bibfield
  {journal} {\bibinfo  {journal} {Phys. Rev.}\ }\textbf {\bibinfo {volume}
  {D90}},\ \bibinfo {pages} {124032} (\bibinfo {year} {2014})},\ \bibinfo
  {note} {[Erratum: Phys. Rev.D94,no.6,069902(2016)]},\ \Eprint
  {http://arxiv.org/abs/1404.3197} {arXiv:1404.3197 [gr-qc]} \BibitemShut
  {NoStop}%
%%CITATION = ARXIV:1404.3197;%%
\bibitem [{\citenamefont {Kamaretsos}\ \emph
  {et~al.}(2012{\natexlab{b}})\citenamefont {Kamaretsos}, \citenamefont
  {Hannam},\ and\ \citenamefont {Sathyaprakash}}]{Kamaretsos:2012bs}%
  \BibitemOpen
  \bibfield  {author} {\bibinfo {author} {\bibfnamefont {I.}~\bibnamefont
  {Kamaretsos}}, \bibinfo {author} {\bibfnamefont {M.}~\bibnamefont {Hannam}},
  \ and\ \bibinfo {author} {\bibfnamefont {B.}~\bibnamefont {Sathyaprakash}},\
  }\href {\doibase 10.1103/PhysRevLett.109.141102} {\bibfield  {journal}
  {\bibinfo  {journal} {Phys. Rev. Lett.}\ }\textbf {\bibinfo {volume} {109}},\
  \bibinfo {pages} {141102} (\bibinfo {year} {2012}{\natexlab{b}})},\ \Eprint
  {http://arxiv.org/abs/1207.0399} {arXiv:1207.0399 [gr-qc]} \BibitemShut
  {NoStop}%
%%CITATION = ARXIV:1207.0399;%%
\bibitem [{\citenamefont {Berti}\ \emph
  {et~al.}(2007{\natexlab{c}})\citenamefont {Berti}, \citenamefont {Cardoso},
  \citenamefont {Gonzalez},\ and\ \citenamefont {Sperhake}}]{Berti:2007dg}%
  \BibitemOpen
  \bibfield  {author} {\bibinfo {author} {\bibfnamefont {E.}~\bibnamefont
  {Berti}}, \bibinfo {author} {\bibfnamefont {V.}~\bibnamefont {Cardoso}},
  \bibinfo {author} {\bibfnamefont {J.~A.}\ \bibnamefont {Gonzalez}}, \ and\
  \bibinfo {author} {\bibfnamefont {U.}~\bibnamefont {Sperhake}},\ }\href
  {\doibase 10.1103/PhysRevD.75.124017} {\bibfield  {journal} {\bibinfo
  {journal} {Phys. Rev.}\ }\textbf {\bibinfo {volume} {D75}},\ \bibinfo {pages}
  {124017} (\bibinfo {year} {2007}{\natexlab{c}})},\ \Eprint
  {http://arxiv.org/abs/gr-qc/0701086} {arXiv:gr-qc/0701086 [gr-qc]}
  \BibitemShut {NoStop}%
%%CITATION = GR-QC/0701086;%%
\bibitem [{\citenamefont {Berti}\ \emph
  {et~al.}(2006{\natexlab{b}})\citenamefont {Berti}, \citenamefont {Cardoso},\
  and\ \citenamefont {Casals}}]{Berti:2005gp}%
  \BibitemOpen
  \bibfield  {author} {\bibinfo {author} {\bibfnamefont {E.}~\bibnamefont
  {Berti}}, \bibinfo {author} {\bibfnamefont {V.}~\bibnamefont {Cardoso}}, \
  and\ \bibinfo {author} {\bibfnamefont {M.}~\bibnamefont {Casals}},\ }\href
  {\doibase 10.1103/PhysRevD.73.109902, 10.1103/PhysRevD.73.024013} {\bibfield
  {journal} {\bibinfo  {journal} {Phys. Rev.}\ }\textbf {\bibinfo {volume}
  {D73}},\ \bibinfo {pages} {024013} (\bibinfo {year} {2006}{\natexlab{b}})},\
  \bibinfo {note} {[Erratum: Phys. Rev.D73,109902(2006)]},\ \Eprint
  {http://arxiv.org/abs/gr-qc/0511111} {arXiv:gr-qc/0511111 [gr-qc]}
  \BibitemShut {NoStop}%
%%CITATION = GR-QC/0511111;%%
\bibitem [{\citenamefont {Kelly}\ and\ \citenamefont
  {Baker}(2013)}]{Kelly:2012nd}%
  \BibitemOpen
  \bibfield  {author} {\bibinfo {author} {\bibfnamefont {B.~J.}\ \bibnamefont
  {Kelly}}\ and\ \bibinfo {author} {\bibfnamefont {J.~G.}\ \bibnamefont
  {Baker}},\ }\href {\doibase 10.1103/PhysRevD.87.084004} {\bibfield  {journal}
  {\bibinfo  {journal} {Phys. Rev.}\ }\textbf {\bibinfo {volume} {D87}},\
  \bibinfo {pages} {084004} (\bibinfo {year} {2013})},\ \Eprint
  {http://arxiv.org/abs/1212.5553} {arXiv:1212.5553 [gr-qc]} \BibitemShut
  {NoStop}%
%%CITATION = ARXIV:1212.5553;%%
\bibitem [{\citenamefont {Berti}\ and\ \citenamefont
  {Klein}(2014)}]{Berti:2014fga}%
  \BibitemOpen
  \bibfield  {author} {\bibinfo {author} {\bibfnamefont {E.}~\bibnamefont
  {Berti}}\ and\ \bibinfo {author} {\bibfnamefont {A.}~\bibnamefont {Klein}},\
  }\href {\doibase 10.1103/PhysRevD.90.064012} {\bibfield  {journal} {\bibinfo
  {journal} {Phys. Rev.}\ }\textbf {\bibinfo {volume} {D90}},\ \bibinfo {pages}
  {064012} (\bibinfo {year} {2014})},\ \Eprint {http://arxiv.org/abs/1408.1860}
  {arXiv:1408.1860 [gr-qc]} \BibitemShut {NoStop}%
%%CITATION = ARXIV:1408.1860;%%
\bibitem [{\citenamefont {Amaro-Seoane}\ \emph {et~al.}(2013)\citenamefont
  {Amaro-Seoane} \emph {et~al.}}]{AmaroSeoane:2012km}%
  \BibitemOpen
  \bibfield  {author} {\bibinfo {author} {\bibfnamefont {P.}~\bibnamefont
  {Amaro-Seoane}} \emph {et~al.},\ }\href@noop {} {\bibfield  {journal}
  {\bibinfo  {journal} {GW Notes}\ }\textbf {\bibinfo {volume} {6}},\ \bibinfo
  {pages} {4} (\bibinfo {year} {2013})},\ \Eprint
  {http://arxiv.org/abs/1201.3621} {arXiv:1201.3621 [astro-ph.CO]} \BibitemShut
  {NoStop}%
%%CITATION = ARXIV:1201.3621;%%
\bibitem [{\citenamefont {Audley}\ \emph {et~al.}(2017)\citenamefont {Audley}
  \emph {et~al.}}]{Audley:2017drz}%
  \BibitemOpen
  \bibfield  {author} {\bibinfo {author} {\bibfnamefont {H.}~\bibnamefont
  {Audley}} \emph {et~al.},\ }\href@noop {} {\  (\bibinfo {year} {2017})},\
  \Eprint {http://arxiv.org/abs/1702.00786} {arXiv:1702.00786 [astro-ph.IM]}
  \BibitemShut {NoStop}%
%%CITATION = ARXIV:1702.00786;%%
\bibitem [{\citenamefont {Jani}\ \emph {et~al.}(2016)\citenamefont {Jani},
  \citenamefont {Healy}, \citenamefont {Clark}, \citenamefont {London},
  \citenamefont {Laguna},\ and\ \citenamefont {Shoemaker}}]{Jani:2016wkt}%
  \BibitemOpen
  \bibfield  {author} {\bibinfo {author} {\bibfnamefont {K.}~\bibnamefont
  {Jani}}, \bibinfo {author} {\bibfnamefont {J.}~\bibnamefont {Healy}},
  \bibinfo {author} {\bibfnamefont {J.~A.}\ \bibnamefont {Clark}}, \bibinfo
  {author} {\bibfnamefont {L.}~\bibnamefont {London}}, \bibinfo {author}
  {\bibfnamefont {P.}~\bibnamefont {Laguna}}, \ and\ \bibinfo {author}
  {\bibfnamefont {D.}~\bibnamefont {Shoemaker}},\ }\href {\doibase
  10.1088/0264-9381/33/20/204001} {\bibfield  {journal} {\bibinfo  {journal}
  {Class. Quant. Grav.}\ }\textbf {\bibinfo {volume} {33}},\ \bibinfo {pages}
  {204001} (\bibinfo {year} {2016})},\ \Eprint
  {http://arxiv.org/abs/1605.03204} {arXiv:1605.03204 [gr-qc]} \BibitemShut
  {NoStop}%
%%CITATION = ARXIV:1605.03204;%%
\bibitem [{\citenamefont {Nollert}(2000{\natexlab{b}})}]{Nollert}%
  \BibitemOpen
  \bibfield  {author} {\bibinfo {author} {\bibfnamefont {H.}~\bibnamefont
  {Nollert}},\ }\href@noop {} {\emph {\bibinfo {title} {{Characteristic
  Oscillations of Black Holes and Neutron Stars: From Mathematical Background
  to Astrophysical Applications}}}}\ (\bibinfo  {publisher} {unpublished
  Habilitationsschrift},\ \bibinfo {year} {2000})\BibitemShut {NoStop}%
\bibitem [{\citenamefont {Kesden}\ \emph {et~al.}(2010)\citenamefont {Kesden},
  \citenamefont {Sperhake},\ and\ \citenamefont {Berti}}]{Kesden:2010yp}%
  \BibitemOpen
  \bibfield  {author} {\bibinfo {author} {\bibfnamefont {M.}~\bibnamefont
  {Kesden}}, \bibinfo {author} {\bibfnamefont {U.}~\bibnamefont {Sperhake}}, \
  and\ \bibinfo {author} {\bibfnamefont {E.}~\bibnamefont {Berti}},\ }\href
  {\doibase 10.1103/PhysRevD.81.084054} {\bibfield  {journal} {\bibinfo
  {journal} {Phys. Rev.}\ }\textbf {\bibinfo {volume} {D81}},\ \bibinfo {pages}
  {084054} (\bibinfo {year} {2010})},\ \Eprint {http://arxiv.org/abs/1002.2643}
  {arXiv:1002.2643 [astro-ph.GA]} \BibitemShut {NoStop}%
%%CITATION = ARXIV:1002.2643;%%
\bibitem [{\citenamefont {Kesden}\ \emph {et~al.}(2015)\citenamefont {Kesden},
  \citenamefont {Gerosa}, \citenamefont {O'Shaughnessy}, \citenamefont
  {Berti},\ and\ \citenamefont {Sperhake}}]{Kesden:2014sla}%
  \BibitemOpen
  \bibfield  {author} {\bibinfo {author} {\bibfnamefont {M.}~\bibnamefont
  {Kesden}}, \bibinfo {author} {\bibfnamefont {D.}~\bibnamefont {Gerosa}},
  \bibinfo {author} {\bibfnamefont {R.}~\bibnamefont {O'Shaughnessy}}, \bibinfo
  {author} {\bibfnamefont {E.}~\bibnamefont {Berti}}, \ and\ \bibinfo {author}
  {\bibfnamefont {U.}~\bibnamefont {Sperhake}},\ }\href {\doibase
  10.1103/PhysRevLett.114.081103} {\bibfield  {journal} {\bibinfo  {journal}
  {Phys. Rev. Lett.}\ }\textbf {\bibinfo {volume} {114}},\ \bibinfo {pages}
  {081103} (\bibinfo {year} {2015})},\ \Eprint {http://arxiv.org/abs/1411.0674}
  {arXiv:1411.0674 [gr-qc]} \BibitemShut {NoStop}%
%%CITATION = ARXIV:1411.0674;%%
\bibitem [{\citenamefont {Gerosa}\ \emph
  {et~al.}(2015{\natexlab{a}})\citenamefont {Gerosa}, \citenamefont {Kesden},
  \citenamefont {Sperhake}, \citenamefont {Berti},\ and\ \citenamefont
  {O'Shaughnessy}}]{Gerosa:2015tea}%
  \BibitemOpen
  \bibfield  {author} {\bibinfo {author} {\bibfnamefont {D.}~\bibnamefont
  {Gerosa}}, \bibinfo {author} {\bibfnamefont {M.}~\bibnamefont {Kesden}},
  \bibinfo {author} {\bibfnamefont {U.}~\bibnamefont {Sperhake}}, \bibinfo
  {author} {\bibfnamefont {E.}~\bibnamefont {Berti}}, \ and\ \bibinfo {author}
  {\bibfnamefont {R.}~\bibnamefont {O'Shaughnessy}},\ }\href {\doibase
  10.1103/PhysRevD.92.064016} {\bibfield  {journal} {\bibinfo  {journal} {Phys.
  Rev.}\ }\textbf {\bibinfo {volume} {D92}},\ \bibinfo {pages} {064016}
  (\bibinfo {year} {2015}{\natexlab{a}})},\ \Eprint
  {http://arxiv.org/abs/1506.03492} {arXiv:1506.03492 [gr-qc]} \BibitemShut
  {NoStop}%
%%CITATION = ARXIV:1506.03492;%%
\bibitem [{\citenamefont {Gerosa}\ \emph
  {et~al.}(2015{\natexlab{b}})\citenamefont {Gerosa}, \citenamefont {Kesden},
  \citenamefont {O'Shaughnessy}, \citenamefont {Klein}, \citenamefont {Berti},
  \citenamefont {Sperhake},\ and\ \citenamefont {Trifirò}}]{Gerosa:2015hba}%
  \BibitemOpen
  \bibfield  {author} {\bibinfo {author} {\bibfnamefont {D.}~\bibnamefont
  {Gerosa}}, \bibinfo {author} {\bibfnamefont {M.}~\bibnamefont {Kesden}},
  \bibinfo {author} {\bibfnamefont {R.}~\bibnamefont {O'Shaughnessy}}, \bibinfo
  {author} {\bibfnamefont {A.}~\bibnamefont {Klein}}, \bibinfo {author}
  {\bibfnamefont {E.}~\bibnamefont {Berti}}, \bibinfo {author} {\bibfnamefont
  {U.}~\bibnamefont {Sperhake}}, \ and\ \bibinfo {author} {\bibfnamefont
  {D.}~\bibnamefont {Trifirò}},\ }\href {\doibase
  10.1103/PhysRevLett.115.141102} {\bibfield  {journal} {\bibinfo  {journal}
  {Phys. Rev. Lett.}\ }\textbf {\bibinfo {volume} {115}},\ \bibinfo {pages}
  {141102} (\bibinfo {year} {2015}{\natexlab{b}})},\ \Eprint
  {http://arxiv.org/abs/1506.09116} {arXiv:1506.09116 [gr-qc]} \BibitemShut
  {NoStop}%
%%CITATION = ARXIV:1506.09116;%%
\bibitem [{\citenamefont {Barausse}\ and\ \citenamefont
  {Buonanno}(2010)}]{Barausse:2009xi}%
  \BibitemOpen
  \bibfield  {author} {\bibinfo {author} {\bibfnamefont {E.}~\bibnamefont
  {Barausse}}\ and\ \bibinfo {author} {\bibfnamefont {A.}~\bibnamefont
  {Buonanno}},\ }\href {\doibase 10.1103/PhysRevD.81.084024} {\bibfield
  {journal} {\bibinfo  {journal} {Phys. Rev.}\ }\textbf {\bibinfo {volume}
  {D81}},\ \bibinfo {pages} {084024} (\bibinfo {year} {2010})},\ \Eprint
  {http://arxiv.org/abs/0912.3517} {arXiv:0912.3517 [gr-qc]} \BibitemShut
  {NoStop}%
%%CITATION = ARXIV:0912.3517;%%
\bibitem [{\citenamefont {Pan}\ \emph {et~al.}(2011)\citenamefont {Pan},
  \citenamefont {Buonanno}, \citenamefont {Fujita}, \citenamefont {Racine},\
  and\ \citenamefont {Tagoshi}}]{Pan:2010hz}%
  \BibitemOpen
  \bibfield  {author} {\bibinfo {author} {\bibfnamefont {Y.}~\bibnamefont
  {Pan}}, \bibinfo {author} {\bibfnamefont {A.}~\bibnamefont {Buonanno}},
  \bibinfo {author} {\bibfnamefont {R.}~\bibnamefont {Fujita}}, \bibinfo
  {author} {\bibfnamefont {E.}~\bibnamefont {Racine}}, \ and\ \bibinfo {author}
  {\bibfnamefont {H.}~\bibnamefont {Tagoshi}},\ }\href {\doibase
  10.1103/PhysRevD.83.064003, 10.1103/PhysRevD.87.109901} {\bibfield  {journal}
  {\bibinfo  {journal} {Phys. Rev.}\ }\textbf {\bibinfo {volume} {D83}},\
  \bibinfo {pages} {064003} (\bibinfo {year} {2011})},\ \bibinfo {note}
  {[Erratum: Phys. Rev.D87,no.10,109901(2013)]},\ \Eprint
  {http://arxiv.org/abs/1006.0431} {arXiv:1006.0431 [gr-qc]} \BibitemShut
  {NoStop}%
%%CITATION = ARXIV:1006.0431;%%
\bibitem [{\citenamefont {Barausse}\ \emph {et~al.}(2014)\citenamefont
  {Barausse}, \citenamefont {Cardoso},\ and\ \citenamefont
  {Pani}}]{Barausse:2014tra}%
  \BibitemOpen
  \bibfield  {author} {\bibinfo {author} {\bibfnamefont {E.}~\bibnamefont
  {Barausse}}, \bibinfo {author} {\bibfnamefont {V.}~\bibnamefont {Cardoso}}, \
  and\ \bibinfo {author} {\bibfnamefont {P.}~\bibnamefont {Pani}},\ }\href
  {\doibase 10.1103/PhysRevD.89.104059} {\bibfield  {journal} {\bibinfo
  {journal} {Phys. Rev.}\ }\textbf {\bibinfo {volume} {D89}},\ \bibinfo {pages}
  {104059} (\bibinfo {year} {2014})},\ \Eprint {http://arxiv.org/abs/1404.7149}
  {arXiv:1404.7149 [gr-qc]} \BibitemShut {NoStop}%
%%CITATION = ARXIV:1404.7149;%%
\bibitem [{\citenamefont {Gossan}\ \emph {et~al.}(2012)\citenamefont {Gossan},
  \citenamefont {Veitch},\ and\ \citenamefont {Sathyaprakash}}]{Gossan:2011ha}%
  \BibitemOpen
  \bibfield  {author} {\bibinfo {author} {\bibfnamefont {S.}~\bibnamefont
  {Gossan}}, \bibinfo {author} {\bibfnamefont {J.}~\bibnamefont {Veitch}}, \
  and\ \bibinfo {author} {\bibfnamefont {B.~S.}\ \bibnamefont
  {Sathyaprakash}},\ }\href {\doibase 10.1103/PhysRevD.85.124056} {\bibfield
  {journal} {\bibinfo  {journal} {Phys. Rev.}\ }\textbf {\bibinfo {volume}
  {D85}},\ \bibinfo {pages} {124056} (\bibinfo {year} {2012})},\ \Eprint
  {http://arxiv.org/abs/1111.5819} {arXiv:1111.5819 [gr-qc]} \BibitemShut
  {NoStop}%
%%CITATION = ARXIV:1111.5819;%%
\bibitem [{\citenamefont {Ghosh}\ \emph {et~al.}(2016)\citenamefont {Ghosh}
  \emph {et~al.}}]{Ghosh:2016qgn}%
  \BibitemOpen
  \bibfield  {author} {\bibinfo {author} {\bibfnamefont {A.}~\bibnamefont
  {Ghosh}} \emph {et~al.},\ }\href {\doibase 10.1103/PhysRevD.94.021101}
  {\bibfield  {journal} {\bibinfo  {journal} {Phys. Rev.}\ }\textbf {\bibinfo
  {volume} {D94}},\ \bibinfo {pages} {021101} (\bibinfo {year} {2016})},\
  \Eprint {http://arxiv.org/abs/1602.02453} {arXiv:1602.02453 [gr-qc]}
  \BibitemShut {NoStop}%
%%CITATION = ARXIV:1602.02453;%%
\end{thebibliography}


\end{document}